\documentclass{jfm}
\usepackage{graphicx}
\usepackage{csquotes}
\usepackage{bigints}
\usepackage{epstopdf, epsfig}
\usepackage{amsmath,amsfonts,amssymb,bm}
\usepackage{mathptmx}
\usepackage{newtxtext}
\usepackage{newtxmath}
\usepackage{bm}
\usepackage[hidelinks]{hyperref}

\usepackage[dvipsnames]{xcolor}
\usepackage{todonotes}
\usepackage{siunitx}
\usepackage{cleveref}

\DeclareSIUnit{\days}{days}
\newcommand{\Ro}{\mathrm{Ro}}
\newcommand{\Bu}{\mathrm{Bu}}
\renewcommand{\vec}[1]{\boldsymbol{#1}}

\renewcommand{\d}{\mathrm{d}}
\newcommand{\pp}[2]{\frac{\partial #1}{\partial #2}}
\newcommand{\dd}[2]{\frac{\d #1}{\d #2}}
\newcommand{\vcdot}{\boldsymbol{\cdot}}
\DeclareMathOperator{\J}{J}

\shorttitle{Regimes of near-inertial wave dynamics}
\shortauthor{S. Conn, J. Callies and A. Lawrence}

\title{Regimes of near-inertial wave dynamics}

\author{Scott Conn\aff{1}
  \corresp{\email{sconn@caltech.edu}},
  Jörn Callies\aff{1} and
  Albion Lawrence\aff{1,2}}

\affiliation{\aff{1}California Institute of Technology, Pasadena, CA 91125, USA \aff{2}Brandeis University, Waltham, MA 02453, USA}

\begin{document}

\maketitle

\begin{abstract}
  When atmospheric storms pass over the ocean, they resonantly force near-inertial waves (NIWs); internal waves with a frequency close to the local Coriolis frequency~$f$. It has long been recognised that the evolution of NIWs is modulated by the ocean’s mesoscale eddy field. This can result in NIWs being concentrated into anticyclones and provide an efficient pathway for their propagation to depth. Whether mesoscale eddies are effective at modulating the behaviour of NIWs depends on the wave dispersiveness $\varepsilon^2 = f\lambda^2/\Psi$, where $\lambda$ is the deformation radius and $\Psi$ is a scaling for the eddy streamfunction. If $\varepsilon\gg1$, NIWs are strongly dispersive, and the waves are only weakly affected by the eddies. We calculate the perturbations away from a uniform wave field and the frequency shift away from~$f$. If $\varepsilon\ll1$, NIWs are weakly dispersive, and the wave evolution is strongly modulated by the eddy field. In this weakly dispersive limit, ray-tracing emerges as a valid description of the NIW evolution even if the large-scale atmospheric forcing apparently violates the requisite assumption of a scale separation between the waves and the eddies. The large-scale forcing excites many wave modes, each of which varies on a short spatial scale and is amenable to asymptotic analysis analogous to the semi-classical analysis of quantum systems. The strong modulation of weakly dispersive NIWs by eddies has the potential to modulate the energy input into NIWs from the wind, but under oceanic conditions, this effect should be small.
\end{abstract}


\section{Introduction}

Near-inertial waves (NIWs) play an important role in the global climate system. Being associated with strong vertical shears, they are prone to shear instabilities which are an important driver of upper ocean mixing \citep[for a review, see][]{alford2016}. As such, the generation of NIWs is one of the primary mechanisms by which atmospheric storms induce a deepening of the surface mixed layer. This deepening requires mixing with water from below, implicating NIWs in the surface ocean heat budget \citep{jochum2013}. In the interior of the ocean, NIWs make up a major fraction of the internal wave kinetic energy \citep{ferrari2009,alford2016}, and it has been hypothesised that NIW kinetic energy may provide a source of mixing in the deep ocean \citep{munk1998abyssal}. NIWs might also extract energy from mesoscale eddies \citep{xie2015,rocha2018} and hence play a role in the mesoscale energy budget.

\textit{In-situ} observations of NIWs usually lack significant spatial resolution. The spatial structure of NIWs can generally only be resolved through dedicated field campaigns. Despite this, it has become clear that NIW evolution can be strongly modulated by the presence of mesoscale eddies \citep{thomas2020,conn2024interpreting}. Given the sparsity of NIW observations, theoretical progress has been important in understanding the dynamics of NIWs in the upper ocean.

Early work on NIW--eddy interactions was based on ray-tracing theory. \citet{kunze1985} derived a dispersion relation for NIWs in the presence of a geostrophic background flow. Throughout this paper, we will make the assumption of a barotropic background flow. The ray-tracing equations for a single baroclinic mode propagating through such a background field are
\begin{equation}
  \dd{\vec{x}}{\tau} = \pp{\omega}{\vec{k}}, \qquad \dd{\vec{k}}{\tau} = -\pp{\omega}{\vec{x}}, \qquad \omega =\frac{f\lambda^2|\vec{k}|^2}{2} + \vec{u}\vcdot\vec{k}+ \frac{\zeta}{2},
  \label{eq:Kunze_disp}
\end{equation}
where $\vec{x}=(x,y)$ is the ray position, $\tau$ is time, $\vec{k}$ is the horizontal wavevector, $\vec{u}$~is the background velocity, $\zeta=\partial_xv-\partial_yu$ is the background vorticity, and $\lambda$ is the deformation radius. Here and throughout the rest of this paper, $\omega$ refers to the frequency shift of an NIW away from the local $f$ such that the true frequency is $f+\omega$. 
Based on these equations, \citet{kunze1985} argued that NIWs would be trapped in regions of anticyclonic vorticity where the effective frequency is less than the local~$f$. This trapping arises from the refraction of rays by the background vorticity, i.e., from changes in the wavenumber vector due to spatial gradients of the $\frac{\zeta}{2}$ term in the dispersion relation. Concentration of NIW energy into anticyclones has indeed been observed in the ocean \citep[e.g.,][]{perkins1976observed,kunze1984observations,thomas2020,yu2022}

Ray-tracing is based on the assumption that the NIWs are propagating through a slowly varying medium. This means that the horizontal scale of the waves has to be much smaller than the scale of the background mesoscale eddy field. \citet[][from hereon YBJ]{young1997} criticised this spatial scale assumption based on the argument that NIWs are forced by large-scale storms and so, at least initially, the waves have a much larger scale than mesoscale eddies. As a remedy, YBJ developed a theory of NIW--eddy interactions that does not rely on the assumption of a spatial scale separation. This was also partly motivated by a desire to explain observations from the Ocean Storms Experiment \citep{dasaro1995a}. This field campaign studied the evolution of NIWs in the wake of a large storm in the North Pacific. A key result of this study was that the effect of the mesoscale vorticity on the wave evolution was in clear contradiction with predictions from ray-tracing \citep{dasaro1995c}. 

The YBJ equation describes the evolution of NIWs in the presence of a prescribed geostrophic eddy field while only assuming a temporal scale separation between the inertial period and the characteristic time scale of the eddies. For the barotropic background flow considered throughout this paper, the wave evolution can be split into baroclinic modes that do not interact, so we consider a single baroclinic mode with NIW velocity $[u_w(x,y,t),v_w(x,y,t)]g(z)$, where $g(z)$ is the baroclinic mode structure. The YBJ equation is cast in terms of the variable $\phi=(u_w+iv_w)e^{ift}$, where the factor $e^{ift}$ removes oscillations at the inertial frequency and leaves $\phi$ to describe the slow evolution of the envelope that modulates the NIWs. The YBJ equation, restricted to a single mode propagating through a barotropic background flow, is then given by
\begin{equation}
    \pp{\phi}{t} + \J(\psi,\phi) + \frac{i\zeta}{2}\phi - \frac{if\lambda^2}{2}\nabla^2\phi = 0\label{eq:3DYBJ},
\end{equation}
where $\psi$ is the background streamfunction, $\zeta=\nabla^2\psi$ is the background vorticity, and $\J(a,b) = \partial_x a \, \partial_y b - \partial_y a \, \partial_x b$ is the Jacobian operator. The second term describes advection of the NIW field by the background flow. The third term is known as the $\zeta$-refraction term and describes refraction of the NIW field by the background vorticity. This term is necessary to obtain concentration of NIWs into regions of anticyclonic vorticity. The last term is responsible for wave dispersion. Here and throughout this paper, we set the meridional gradient of planetary vorticity $\beta=0$. The YBJ equation can be modified to include $\beta$ by replacing $\zeta/2$ with $\zeta/2+\beta y$ in the refraction term. The $\beta$-effect has been proposed to explain the observed equatorward propagation of NIWs in the ocean \citep{anderson1979beta,garrett2001near,yu2022}. Over short enough scales, the $\beta y$ term will only provide a small correction to $\frac{\zeta}{2}$ and hence we ignore it.

Despite both ray-tracing and the YBJ equation being used in the NIW literature, it remains unclear how they relate to each other. Ray-tracing has been one of the most widely used tools to interpret observations of NIWs. It has revealed aspects of NIW dynamics such as trapping in anticyclones along with an associated propagation to depth \citep{jaimes2010near}, stalling in cyclones \citep{oey2008stalling}, and the interplay between NIWs and turbulent dissipation \citep{kunze1995energy,essink2022}. Non-standard propagation patterns of NIWs in observations have also been explained using ray-tracing \citep[e.g.,][]{byun2010observation,chen2013observed}. The YBJ equation has been used primarily as a tool in theoretical and numerical studies, although there has been some attempt to make connections with observations. \cite{thomas2020} calculated the NIW wavevector using an expression based on the YBJ equation. The predictions from YBJ were broadly in agreement with observations. \cite{conn2024interpreting} directly used the YBJ equation to interpret NIW observations on a mooring array, showing that it successfully captured the amplitude and phase evolution, including differences across the mooring caused by mesoscale vorticity gradients. Comparing the results of these disparate studies is complicated by the different methods used. Having a better understanding of the relationship between ray-tracing and YBJ would facilitate the comparison of these results. 

Furthermore, observations reveal a varied picture of the importance of the mesoscale vorticity on NIW evolution. During the Ocean Storms Experiment, mesoscale eddies had a muted impact on the NIW field \citep{dasaro1995c}, whereas other observational studies found a strong imprint of mesoscale eddies onto the NIW field. For example, \citet{thomas2020} demonstrated that the evolution of the NIW wavevector was driven by gradients in the mesoscale vorticity during the NISKINe experiment in the North Atlantic. Extending the original argument by YBJ, \cite{thomas2024why} argued that these differences in the impact of mesoscale vorticity could be explained primarily by differences in the strength of wave dispersion. The stronger dispersion in the Ocean Storms Experiment, they argued, was the result of the forcing projecting onto lower baroclinic modes, a stronger stratification, and weaker eddies. As a result, the effect of refraction by mesoscale vorticity was suppressed in the Ocean Storms Experiment, whereas it was more pronounced in NISKINe.

In this paper we clarify the following question about NIW dynamics: how does the ray-tracing approach relate to YBJ dynamics? Given the widespread use of ray-tracing in the literature, we aim to understand the conditions under which results from ray-tracing are accurate. To this end, we consider the YBJ equation in both a strong and weak-dispersion regime. We begin by providing a simplified treatment of the strong-dispersion regime. Next, we show that the ray-tracing equations emerge asymptotically from the YBJ equation in the limit of weak dispersion. Finally, we consider how these regimes might modulate the energy injection into the NIW band by the winds, finding that such a modulation is likely weak under oceanic conditions.

\section{The YBJ equation}

\subsection{Decomposition into horizontal modes}
\begin{figure}
  \centering
  \includegraphics{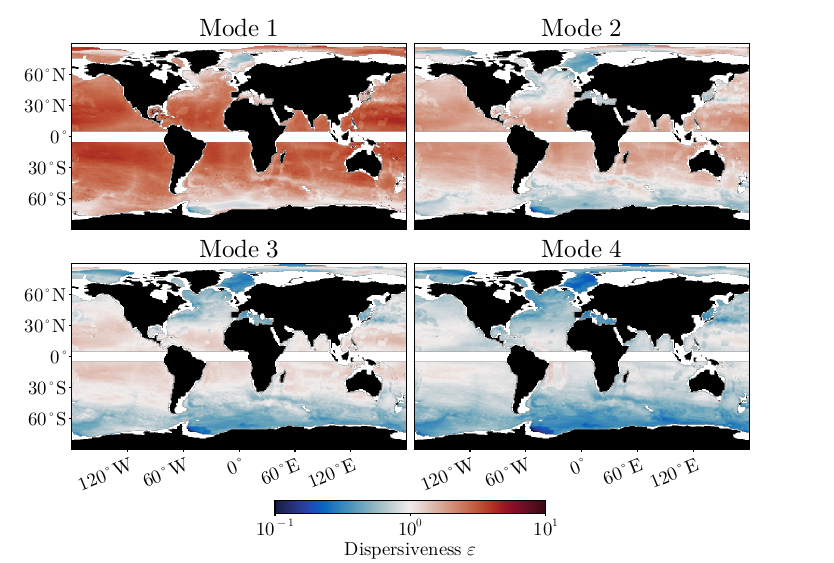}
  \caption{Dispersiveness parameter $\varepsilon = f \lambda^2 / \Psi$ plotted throughout the ocean for the first four baroclinic modes, with the deformation radius~$\lambda$ estimated from hydrography and the streamfunction magnitude~$\Psi$ from altimetry. The equatorial band is blocked out because the mean flow amplitude cannot be estimated with confidence there.}
  \label{fig:epsilon}
\end{figure}

We begin by non-dimensionalising the YBJ equation. Given the scalings $x,y\sim L$, $\psi\sim\Psi$ and $t\sim L^2/\Psi$, we obtain the following non-dimensional form of the YBJ equation:
\begin{equation}
  \pp{\phi}{t} + \J(\psi,\phi) + \frac{i\zeta}{2}\phi - \frac{i\varepsilon^2}{2}\nabla^2\phi = 0,
  \label{eq:2DYBJ}
\end{equation}
where $\varepsilon^2 = f\lambda^2/\Psi$ is the wave dispersivity. For readers familiar with \citet{young1997}, our $\varepsilon^2$ is equivalent to their $\Upsilon^{-1}$. We remind the reader that we have assumed a single baroclinic mode, but $\varepsilon$ does vary among baroclinic modes through $\lambda$. The parameter $\varepsilon$ also varies spatially throughout the ocean (Fig.~\ref{fig:epsilon}). We calculate $\varepsilon$ for the first four baroclinic modes from observations as described in Appendix~\ref{app_eps}. With the exception of the high latitudes, the first and second baroclinic modes are almost entirely in the strongly dispersive regime ($\varepsilon>1$). Higher baroclinic modes tend to be more weakly dispersive with $\varepsilon<1$ almost everywhere for mode~4. For a given baroclinic mode, low-latitude regions tend to be more strongly dispersive while higher latitudes and western boundary currents are more weakly dispersive.

Note that \eqref{eq:2DYBJ} is a Schr\"odinger equation. This parallel is made clear if we write \eqref{eq:2DYBJ} as
\begin{equation}
    i\pp{\phi}{t} = H\phi, \qquad H = -\frac{\varepsilon^2}{2}\nabla^2  - i \J(\psi, \, \cdot \,) + \frac{\zeta}{2} . \label{eq:eval_2D}
\end{equation}
The operator $H$ is known as the Hamiltonian operator. While the presence of first derivatives in the Hamiltonian may be unfamiliar to some, such terms arise in quantum mechanics when describing a charged particle in a magnetic field. In the YBJ equation, these first derivative terms arise due to advection. This analogy to quantum mechanics was pointed out by \cite{balmforth1998}; we will here exploit it extensively.

This operator $H$ is Hermitian and so it has real eigenvalues. Let $\vec{\mu}$ label the eigenmodes $\phi_{\vec{\mu}}(x,y)$ and associated eigenvalues $\omega_{\vec{\mu}}$ of the operator $H$,
\begin{equation}
  H \phi_{\vec{\mu}} = \omega_{\vec{\mu}} \phi_{\vec{\mu}}.
  \label{eqn:generaleig}
\end{equation}
We will employ two-component vectors~$\vec{\mu}$ to label the two-dimensional modes. The field $\phi$ can then be expanded in these horizontal eigenmodes:
\begin{equation}
    \phi(x,y,t) = \sum_{\vec{\mu}} a_{\vec{\mu}}(t)\phi_{\vec{\mu}}(x,y),
\end{equation}
where $a_{\vec{\mu}}(t)$ is the projection of $\phi$ onto the eigenmode~$\phi_{\vec{\mu}}$. The coefficients $a_{\vec{\mu}}(t)$ then evolve according to
\begin{equation}
  \dd{a_{\vec{\mu}}}{t} = -i\omega_{\vec{\mu}}a_{\vec{\mu}}, \qquad \text{so} \qquad a_{\vec{\mu}}(t)=a_{\vec{\mu}}(0)e^{-i\omega_{\vec{\mu}}t}.
\end{equation}
Therefore, the eigenvalue represents the frequency shift of the mode away from $f$. Furthermore, because the eigenvalues are real, the total kinetic energy of the waves is conserved.

\subsection{Numerical calculation of eigenvalues and eigenmodes}

For most choices of the background flow $\psi$, analytical solutions for the eigenfunctions of $H$ do not exist and numerical solutions are required. Solving the eigenvalue equation numerically requires us to discretise the operator $H$. The discrete eigenfunction is expressed as a vector, and the problem reduces to finding the eigenvalues of a finite matrix. The operator $H$ is Hermitian and so it is desirable for any discrete representation of $H$ to also be Hermitian. A standard second-order central finite difference scheme for the Laplacian term preserves this property. More care is required for the advection operator, for which we use the enstrophy-conserving scheme from \citet{arakawa1966computational} to preserve the Hermitian nature of the operator and guarantee that the eigenvalues of the matrix are real. Having real eigenvalues ensures that the conservation of NIW kinetic energy is respected in the discrete system. The exact method of numerically solving the eigenvalue problem is detailed in Appendix~\ref{app_num}.

\section{The strong-dispersion limit}

The limit where $\varepsilon^2\gg 1$ is known as the strong-dispersion limit. YBJ showed that in this limit, the solution to the YBJ equation becomes proportional to the streamfunction $\psi$. They additionally showed that frequency shifts away from $f$ are proportional to the domain-averaged kinetic energy of the mesoscale flow. These same results can be derived by considering the eigenvalue problem posed above. In our framework, we can additionally derive information about the next-order perturbations to the NIW field.

\begin{figure}
    \centering
    \includegraphics{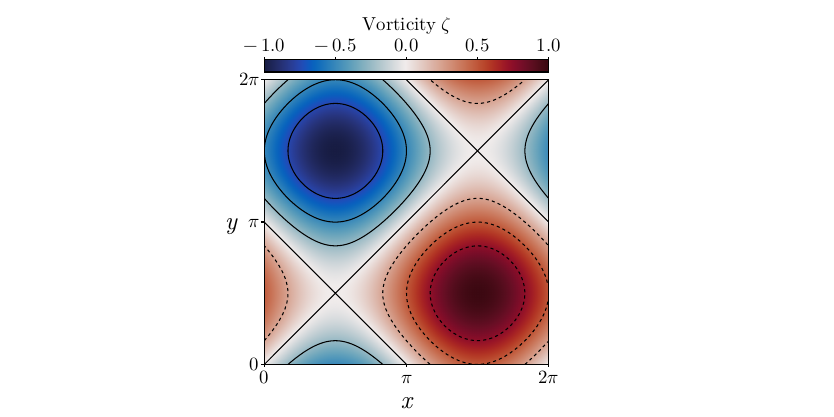}
    \caption{Dipole vorticity with an anticyclone in the upper left corner and a cyclone in the lower right corner. The contours depict the streamfunction with positive values denoted by solid lines and negative values denoted by dashed lines.}
    \label{fig:dipole}
\end{figure}

When $\varepsilon^2$ is large we split the operator $H$ into two parts
\begin{equation}
    H = \varepsilon^2 H^{(0)} + H^{(1)},
\end{equation}
where $H^{(0)}=-\frac{1}{2}\nabla^2$ and $H^{(1)}=\frac{1}{2}\zeta -i\J(\psi,\cdot)$. Because $\varepsilon^2\gg1$, this implies $H^{(1)}$ is a small correction to $\varepsilon^2H^{(0)}$, and perturbation theory can be used to solve this system. We expand both $\phi_{\vec{\mu}}$ and $\omega_{\vec{\mu}}$ in powers of $\varepsilon^{-2}$:
\begin{equation}
  \phi_{\vec{\mu}} = \sum_{n = 0}^\infty \varepsilon^{-2n}\phi_{\vec{\mu}}^{(n)}, \qquad \omega_{\vec{\mu}} = \varepsilon^2 \sum_{n = 0}^\infty \varepsilon^{-2n}\omega_{\vec{\mu}}^{(n)}.
\end{equation}
At $\textit{O}(\varepsilon^{2})$ the eigenvalue problem is
\begin{equation}
    H^{(0)}\phi^{(0)}_{\vec{\mu}} = \omega^{(0)}_{\vec{\mu}}\phi^{(0)}_{\vec{\mu}},
\end{equation}
where $\phi_{\vec{\mu}}^{(0)}$ is the eigenfunction of the unperturbed problem with eigenvalue $\omega^{(0)}_{\vec{\mu}}$. We assume the domain is doubly periodic and goes from $0$ to $2\upi$ in $x$ and~$y$. The solution is
\begin{equation}
  \phi_{\vec{\mu}}^{(0)}=e^{i\vec{\mu}\vcdot\vec{x}}, \qquad \omega_{\vec{\mu}}^{(0)}=\frac{|\vec{\mu}|^2}{2}.
  \label{eq:unpert_eigenfunction}
\end{equation}
The components of $\vec{\mu}$ are integers, and the eigenfunctions are plane waves in $x$ and $y$. The use of a periodic domain is intended to represent a local view of an ocean that is filled with a random sea of eddies. For many examples in this paper, we will consider a domain that contains a dipole vortex (figure~\ref{fig:dipole}) given by \citep[cf.,][]{asselin2020refraction}
\begin{equation}
    \psi = \frac{1}{2}\left(\sin{x}-\sin{y}\right).
    \label{eqn:dipole}
\end{equation}
The analysis below is general, however, and can be applied to any doubly period background flow.

NIWs are forced by atmospheric storms which have a much larger horizontal scale than mesoscale eddies, so it can be idealized as a uniform forcing. The projection of a uniform forcing onto a given mode can be found by integrating that mode across the domain. For plane waves, a domain integral will vanish unless $\vec{\mu} = 0$, such that a uniform forcing will only project onto the $\vec{\mu} = 0$ mode in the unperturbed case. We begin by focusing on that case to obtain expressions for the perturbations to its spatial structure as well as its frequency shift. A small part of the forcing, however, projects onto modes with $\vec{\mu} \neq 0$, and we will return to these higher modes below.

The leading-order solution for $\vec{\mu} = 0$ is horizontally uniform and contains no modulation of the waves by the mesoscale eddy field. To obtain this modulation we must go to higher order. At $\textit{O}(\varepsilon^0)$, the eigenvalue problem is
\begin{equation}
    H^{(0)}\phi^{(1)}_0 + H^{(1)}\phi^{(0)}_0 = \omega^{(0)}_0\phi^{(1)}_0 + \omega^{(1)}_0\phi^{(0)}_0.\label{eq:SD_order0}
\end{equation}
From the $\textit{O}(\varepsilon^{-2})$ calculations, we know $\omega_0^{(0)}=0$. Similarly, the advection term in $H^{(1)}$ vanishes when acting on $\phi_0^{(0)}$ because $\J(\psi,\phi_{\vec{\mu}}^{(0)}) = i \vec{u} \vcdot \vec{\mu}$, and the $\textit{O}(\varepsilon^0)$ equation reduces to
\begin{equation}
    -\frac{1}{2}\nabla^2\phi_{0}^{(1)} + \frac{\zeta}{2} = \omega_0^{(1)}.
\end{equation}
The two terms on the left vanish when integrated over the domain, and we conclude that $\omega_0^{(1)} = 0$.

There is, however, a correction to the eigenfunction at this order, determined by
\begin{equation}
    \nabla^2\phi_0^{(1)} = \nabla^2\psi.
\end{equation}
With periodic boundary conditions, the solution to this is
\begin{equation}
    \phi_0^{(1)} = \psi,
\end{equation}
where we have assumed that $\psi$ is defined such that it has zero domain average. This recovers the expression for $\phi$ from YBJ. The structure of the mesoscale eddy field is imprinted onto the waves by the $\varepsilon^{-2}\phi^{(1)}_0$ term. Because the modulation is by the real streamfunction~$\psi$, only the NIW amplitude is modulated by the mesoscale eddies. The NIW field remains in phase across the domain. 

We now also seek the leading-order correction to the eigenvalue, for which we go up another order. The eigenvalue equation at $\textit{O}(\varepsilon^{-2})$ is
\begin{equation}
    H^{(0)}\phi_0^{(2)} + H^{(1)}\phi_0^{(1)} = \omega^{(0)}\phi_0^{(2)} + \omega^{(1)}\phi_0^{(1)} + \omega^{(2)}\phi_0^{(0)}.
    \label{eq:SD_order_2}
\end{equation}
With $\omega_0^{(0)}=\omega_0^{(1)}=0$ and $\J(\psi,\psi)=0$, this simplifies to
\begin{equation}
    -\frac{1}{2}\nabla^2\phi_0^{(2)} + \frac{1}{2} \psi \nabla^2 \psi = \omega_2^{(0)}.
\end{equation}
The first term on the left vanishes under domain integration. Integrating the second term on the left by parts yields
\begin{equation}
  \omega_0^{(2)} = -\frac{1}{2}\frac{\int |\nabla\psi|^2 \, \d^2 \vec{x}}{\int \d^2 \vec{x}}.
  \label{eq:SD_freq_shift}
\end{equation}
The leading-order frequency shift is $\varepsilon^{-2} \omega_0^{(2)}$. Given that $\varepsilon^{-2} \ll 1$, the frequency shift away from~$f$ is suppressed substantially, even compared to the small frequency shift assumed from the outset. Redimensionalising the expression results in
\begin{equation}
    \omega_0^{(2)} = -\frac{1}{2f_0\lambda^2}\frac{\int |\nabla\psi|^2 \, \d^2 \vec{x}}{\int \d^2 \vec{x}}.
\end{equation}
This agrees with the YBJ result for the dispersion relation in the strong-dispersion regime, indicating that the frequency shift is proportional to the kinetic energy of the eddy field.

We now return to the higher modes with~$\vec{\mu}\neq0$. These modes are degenerate to leading order. For example, the modes $(1,0),(-1,0),(0,1)$ and $(0,-1)$ all have $\omega^{(0)}_{\vec{\mu}}=\frac{1}{2}$. We outline the procedure for using degenerate perturbation theory to calculate corrections to the eigenvalues and eigenfunctions for $\vec{\mu}\neq0$ \citep[e.g.,][]{sakurai2020modern}. We again start from the $\textit{O}(\varepsilon^0)$ equation which now reads
\begin{equation}
  H_0\phi_{\vec{\mu}}^{(1)}+H_1\phi_{\vec{\mu}}^{(0)}=\omega_{\vec{\mu}}^{(0)}\phi_{\vec{\mu}}^{(1)} + \omega_{\vec{\mu}}^{(1)}\phi_{\vec{\mu}}^{(0)}.
  \label{eqn:sdo0}
\end{equation}
Multiplying this equation by $\phi_{\vec{\nu}}^{(0)*}$, with both $\vec{\nu}$ and $\vec{\mu}$ labelling one of the modes in the degenerate group, and integrating over the domain results in
\begin{equation}
  \int\phi_{\vec{\nu}}^{(0)*}\left(H_0-\omega_{\vec{\mu}}^{(0)}\right)\phi_{\vec{\mu}}^{(1)} \, \d^2 \vec{x} = \omega_{\vec{\mu}}^{(1)}\int \phi_{\vec{\nu}}^{(0)*}\phi_{\vec{\mu}}^{(0)} \, \d^2 \vec{x} - \int \phi_{\vec{\nu}}^{(0)*} H_1 \phi_{\vec{\mu}}^{(0)} \, \d^2 \vec{x}.
\end{equation}
Using integration by parts, the $H_0$ on the left can be swapped for $\omega_{\vec{\nu}}^{(0)}$. Because the modes are degenerate to this order, the left-hand side vanishes. Furthermore using orthonormality of the eigenfunctions, the corrections to the eigenvalues are determined by
\begin{equation}
  \omega_{\vec{\mu}}^{(1)} \delta_{\vec{\nu}\vec{\mu}} = \frac{1}{4\upi^2} \int \phi_{\vec{\nu}}^{(0)*} H_1 \phi_{\vec{\mu}}^{(0)} \, \d^2 \vec{x}.
    \label{eq:omega_1_degenerate}
\end{equation}
The left-hand side of this equation is diagonal, which demands that we choose the basis set of the degenerate subspace~$\phi_{\vec{\mu}}^{(0)}$ such that it diagonalises the operator~$H_1$. This can be done by choosing an arbitrary basis, such as the one mentioned above, and then diagonalising the matrix with its elements equal to the right-hand side in~\eqref{eq:omega_1_degenerate}. The corresponding eigenfunction corrections can be found by solving the screened Poisson equation obtained from the first-order equation~\eqref{eqn:sdo0}
\begin{equation}
  \left( H_0-\omega_{\vec{\mu}}^{(0)} \right) \phi_{\vec{\mu}}^{(1)} = \left(\omega_{\vec{\mu}}^{(1)}-H_1\right)\phi_{\vec{\mu}}^{(0)},
  \label{eq:poisson}
\end{equation}
where the $\phi_{\vec{\mu}}^{(0)}$ should be in the basis diagonalising~$H_1$. If~$H_1$ identically vanishes in this subspace, the degeneracy must be lifted at the next order, as in the example below. The same procedure applies.

\begin{figure}
    \centering
    \includegraphics{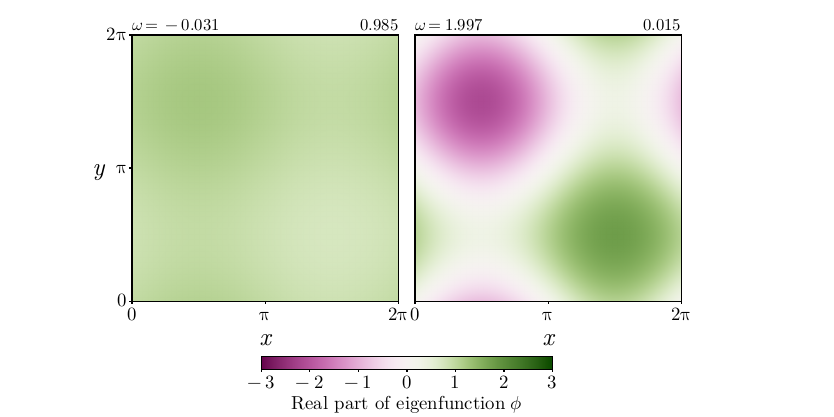}
    \caption{Numerical solution to the eigenvalue problem~\eqref{eqn:generaleig} with $\varepsilon=2$ for the dipole flow. A uniform forcing primarily projects onto the mode shown on the left, with the mode with the second highest projection shown on the right only making up less than 2\% of the energy. The eigenvalues~$\omega$ and projection fractions (of energy) are shown in the panel titles.}
    \label{fig:SD}
\end{figure}

We now consider the specific example of the dipole flow~\eqref{eqn:dipole}. Numerical solutions for $\varepsilon=2$ show that a uniform initial condition projects strongly (98.5\% of the energy) onto the $\phi_0$ mode (figure~\ref{fig:SD}). There is a small but negative frequency shift of $\omega_0 = -0.03104$. This agrees excellently with the predicted frequency shift from \eqref{eq:SD_freq_shift} of $\varepsilon^{-2} \omega_0^{(2)} = \frac{1}{32} = -0.03125$. Additionally, there is weak horizontal structure that aligns with the streamfunction as expected. The root-mean-squared error between the numerical eigenmode and the analytical eigenmode $\phi_0^{(0)} + \varepsilon^{-2} \phi_0^{(1)} = 1 + \varepsilon^{-2} \psi$ is~1\%. The agreement is excellent despite $\varepsilon$ not being particularly large.

For the dipole flow, the right-hand side of \eqref{eq:omega_1_degenerate} is zero for all combinations of basis functions of the $\omega_{\vec{\mu}}^{(0)} = \frac{1}{2}$ subspace. Therefore, there are no first-order frequency shifts, $\omega_{\vec{\mu}}^{(1)} = 0$, and the degeneracy is not lifted at this order. Performing the same procedure that led to~\eqref{eq:omega_1_degenerate} on the second-order equation yields
\begin{equation}
  \omega_{\vec{\mu}}^{(2)} \delta_{\vec{\nu}\vec{\mu}} = \frac{1}{4\upi^2} \int \phi_{\vec{\nu}}^{(0)*} H_1 \phi_{\vec{\mu}}^{(1)} \, \d^2 \vec{x}.
\end{equation}
For our trial basis consisting of the four plane waves, we solve the screened Poisson equation~\eqref{eq:poisson} for the corresponding~$\phi_{\vec{\mu}}^{(1)}$. This is tedious but doable because the right-hand side is just a sum of sines and cosines. The equation for the second-order frequency shift can be diagonalised, and this time the eigenvalues are not zero and the degeneracy is lifted. We find for $\omega_{\vec{\mu}}^{(2)}$ the values $-\frac{1}{96}$, $-\frac{7}{96}$, $-\frac{49}{96}$, and $-\frac{55}{96}$, only the first of which corresponds to an eigenfunction that the forcing projects onto at this order. The leading-order eigenfunction of that mode is $\phi_{\vec{\mu}}^{(0)} = -\psi$ (figure~\ref{fig:SD}). The eigenvalue $\varepsilon^2 \omega_{\mu}^{(0)} + \varepsilon^{-2} \omega_{\vec{\mu}}^{(2)} = 1.99739$ is again in excellent agreement with the numerical eigenvalue of~$1.99729$.

In this regime, horizontal structure in the waves primarily arises due to $\phi_0^{(1)}$ which is suppressed by $\textit{O}(\varepsilon^{-2})$. There is also horizontal structure due to modes with $\vec{\mu}\neq0$ but these are projected onto weakly; the fraction of the variance accounted for by such a mode is $\mathcal{O}(\varepsilon^{-4})$ \citep{sakurai2020modern}.  As such, the wave potential energy, which depends on horizontal gradients in the wave field, is also suppressed. \citet{xie2015} associated the generation of wave potential energy with a sink of the background eddy kinetic energy in a process known as stimulated generation. Given the weak generation of horizontal structure, stimulated generation is weak in the strong-dispersion regime.

\section{The weak-dispersion limit}

The limit $\varepsilon^2 \ll 1$ is known as the weak-dispersion limit. Because $\varepsilon^2$ multiplies the highest-order derivative in the eigenvalue equation, the limit $\varepsilon^2\rightarrow0$ is a singular perturbation. Before addressing the general problem, we build intuition with two simple examples. These examples suggest that there are two classes of modes. One class is characterised by waves that vary slowly along the streamlines of the background flow and more rapidly across streamlines; they are captured by an anisotropic scaling. The other class has even faster variations in both directions and requires an isotropic scaling. We develop a uniformly valid approximation that captures both of these classes.

\subsection{Parallel shear flow}

We begin with an example of a parallel shear flow in which the streamfunction~$\psi$ is a function of~$x$ only. The symmetry in $y$ means the problem reduces to a one-dimensional eigenvalue problem. \cite{balmforth1998} considered this problem for a specific example of a shear flow that can be solved in closed form. \citet{zhang2023scale} considered the limits of strong and weak dispersion for the same mean flow. Here, we address how the weak-dispersion limit can be analysed for a general parallel shear flow and apply the procedure to the example flow from \citet{balmforth1998}.

We assume that the streamfunction~$\psi(x)$ is periodic on the domain $[-\upi,\upi]$. The eigenvalue problem~\eqref{eqn:generaleig} reduces to
\begin{equation}
  -\frac{\varepsilon^2}{2}\nabla^2\phi - iv\pp{\phi}{y} + \frac{\zeta}{2}\phi= \omega\phi,
\end{equation}
where $\zeta = \nabla^2 \psi$ and $v = \partial_x \psi$ are both functions of $x$ only, and we have suppressed the label on the eigenmode. The coefficients are independent of~$y$ which motivates the ansatz $\phi = \Phi(x)e^{imy}$. Given that the domain has width $2\upi$ in $y$, the wavenumber $m$ must be an integer. With this ansatz, we are left with the one-dimensional eigenvalue problem
\begin{equation}
  -\frac{\varepsilon^2}{2}\frac{\d^2 \Phi}{\d x^2} + \left( \frac{\varepsilon^2 m^2}{2} + m v  + \frac{\zeta}{2}\right) \Phi = \omega\Phi
  \label{eq:1D_shear}.
\end{equation}
This is the Schr\"odinger equation of a particle in one-dimensional potential, with the bracketed term playing the role of the potential \citep{balmforth1998}.

As $\varepsilon$ is small, WKB analysis can be used to find approximations to the eigenvalues and eigenfunctions \citep[e.g.,][]{bender1999advanced}. In WKB theory, the field $\Phi$ is expanded as
\begin{equation}
  \Phi(x) = \exp \frac{1}{\delta} \sum_{j = 0}^\infty \delta^j S_j(x),
\end{equation}
where $\delta \ll 1$ is a scaling parameter that we are yet to determine. Substituting this into~\eqref{eq:1D_shear} yields
\begin{equation}
  -\frac{\varepsilon^2}{2}\left[\frac{1}{\delta^2}\left(\sum_{j = 0}^\infty \delta^j\dd{S_j}{x}\right)^2+\frac{1}{\delta} \sum_{j = 0}^\infty \delta^j\dd{^2 S_j}{x^2}\right]+\frac{\varepsilon^2m^2}{2}+mv+\frac{\zeta}{2}=\omega.
\end{equation}
If we assume $m\sim\textit{O}(1)$, both the refraction term and the advection terms are $O(1)$, and they must be balanced by a dispersion term of the same order. Requiring the lowest-order dispersion term to be $\textit{O}(1)$ implies $\delta = \varepsilon$, and the $\textit{O}(1)$ equation becomes
\begin{equation}
    -\frac{1}{2}\left(\dd{S_0}{x}\right)^2 + mv + \frac{\zeta}{2} = \omega.\label{eq:1D_WKB_1}
\end{equation}
By writing $\varepsilon^{-1} S_0'(x) = i k(x)$, this equation is analogous to the dispersion relation~\eqref{eq:Kunze_disp} specialized to this parallel shear flow. The function~$S_0$ is found to be
\begin{equation}
  S_0(x) = \pm \sqrt{2}i \int^x \sqrt{\omega- m v(x') - \frac{\zeta(x')}{2}} \, \d x'
  \label{eq:S0}
\end{equation}
and determines the leading-order phase variations of the solution. One can additionally show \citep[][equation~10.1.12]{bender1999advanced} that the next-order solution is 
\begin{equation}
    S_1(x)=-\frac{1}{4}\ln{\left(\omega-m v-\frac{\zeta}{2}\right)},
\end{equation} 
which determines the leading-order amplitude modulation of the solution.

This asymptotic expansion is valid away from regions where the integrand above is zero. These are known as turning points of the problem, and exist if $\omega < \max(  mv + \zeta/2)$. The associated eigenfunctions are referred to as bound states. Near turning points, $\omega-mv-\zeta/2$ can be approximated by a linear function of $x$, and solutions to~\eqref{eq:1D_WKB_1} are given by Airy functions. The Airy function solutions must be asymptotically matched to the solutions away from the turning points. This yields an integral constraint from which the eigenvalues~$\omega$ can be determined. The problem as formulated above is the classic two-turning point problem and the asymptotic matching procedure is well documented \citep[e.g.,][equation 10.5.6]{bender1999advanced}. The resulting condition for $\omega$, often referred to as a quantisation condition, is 
\begin{equation}
  \frac{\sqrt{2}}{\varepsilon} \int_{x_0}^{x_1} \sqrt{\omega - mv(x) - \frac{\zeta(x)}{2}} \, \d x = \left(n+\frac{1}{2}\right)\upi, \quad \text{with} \quad n = 0, 1, \dots,
\end{equation}
where $x_0$ and $x_1$ are the turning points of the integrand above. The projection of a uniform forcing onto these modes can also be calculated asymptotically. The domain integral of a mode is dominated by contributions from the turning points \citep[e.g.,][equation~10.4.24]{bender1999advanced}.

If $\omega > \max (\zeta/2 + mv)$ then there are no turning points. The corresponding eigenmodes are referred to as free states, and the quantisation condition is replaced by
\begin{equation}
  \frac{\sqrt{2}}{\varepsilon} \int_{-\upi}^{\upi} \sqrt{\omega - mv(x) - \frac{\zeta(x)}{2}} \, \d x = 2n\upi, \quad \text{with} \quad n = 0, 1, \dots
\end{equation}
Note the lack of a half-integer shift that for bound states arises from the Airy behaviour near turning points. The lack of turning points in the free states also means~\eqref{eq:S0} is valid across the entire domain.
Because the eigenfunctions of the free states are oscillatory in the entire domain, a uniform forcing projects only weakly onto them, and we do not discuss them any further. 

Under this scaling, the WKB modes are anisotropic. We assumed $m \sim {O}(1)$, which means that the modes' phase varies in $y$ on a length scale $O(1)$. In contrast, the leading-order phase variations in~$x$ come from $\varepsilon^{-1} S_0$ and therefore occur on a scale~$O(\varepsilon)$. The phase varies slowly along streamlines and rapidly across streamlines. This makes refraction and advection come in at the same order as cross-streamline dispersion. \citet{asselin2020refraction} discussed solutions to the YBJ equation which are aligned with streamlines and for which straining is ineffective. These solutions correspond to our anisotropic modes.

An alternative would be to choose the scaling $m\sim\textit{O}(\varepsilon^{-2}).$ Repeating the WKB ansatz requires a choice of $\delta=\varepsilon^2$ and $\omega\sim\textit{O}(\varepsilon^{-2})$ in order to end up with an equation of a similar form to~\eqref{eq:1D_WKB_1}:
\begin{equation}
  -\frac{1}{2}\left(\dd{S_0}{x}\right)^2 + \frac{\varepsilon^4 m^2}{2} + \varepsilon^2 m v = \varepsilon^2 \omega.
  \label{eq:1D_WKB_2}
\end{equation}
With the scaling given above, each term is $O(1)$. We can solve for $S_0$ and the corresponding quantisation condition for bound modes:
\begin{gather}
  S_0(x) = \pm \sqrt{2}i \varepsilon \int^x\sqrt{\omega-\frac{\varepsilon^2m^2}{2}-mv(x')} \, \d x', \\
  \frac{\sqrt{2}}{\varepsilon}\int_{x_0}^{x_1}\sqrt{\omega-\frac{\varepsilon^2 m^2}{2}-mv} \, \d x = \left(n+\frac{1}{2}\right)\upi.
\end{gather}
These modes are isotropic. The phase variations in~$y$ occur on a scale $O(\varepsilon^2)$, which is the same as in~$x$ because phase variations in~$x$ now come from $\varepsilon^{-2} S_0$. This makes advection and along-streamline dispersion come in at the same order, and it makes refraction negligible.

Despite the different characteristics of the two scalings, they lead to similar quantisation conditions that differ only by what terms are included. We can combine them into a uniformly valid quantisation condition:
\begin{equation}
  \frac{\sqrt{2}}{\varepsilon}\int_{x_0}^{x_1}\sqrt{\omega-\frac{\varepsilon^2m^2}{2} - mv(x) - \frac{\zeta(x)}{2}} \, \d x = \left(n+\frac{1}{2}\right)\upi.
  \label{eq:shear_eval_eqn}
\end{equation}
The ``potential'' governing the wave evolution is therefore
\begin{equation}
  V(x) = \frac{\varepsilon^2m^2}{2} + mv(x) + \frac{\zeta(x)}{2}.
\end{equation}
Under the anisotropic scaling $m \sim O(1)$, the along-streamline dispersion term is suppressed by a factor $\varepsilon^{2}$, leaving the $O(1)$ refraction and advection terms to dominate. Under the isotropic scaling $m \sim O(\varepsilon^{-2})$, the advection and along-streamline dispersion terms are enhanced by a factor~$\varepsilon^{-2}$ and dominate over a now negligible refraction term. In both cases, the general equation is obtained by retaining a term that is of higher order, which is allowed in an asymptotic theory. A uniform forcing only projects onto modes with $m = 0$, so all of the modes projected onto are of the anisotropic variety.

\begin{figure}
  \centering
  \includegraphics{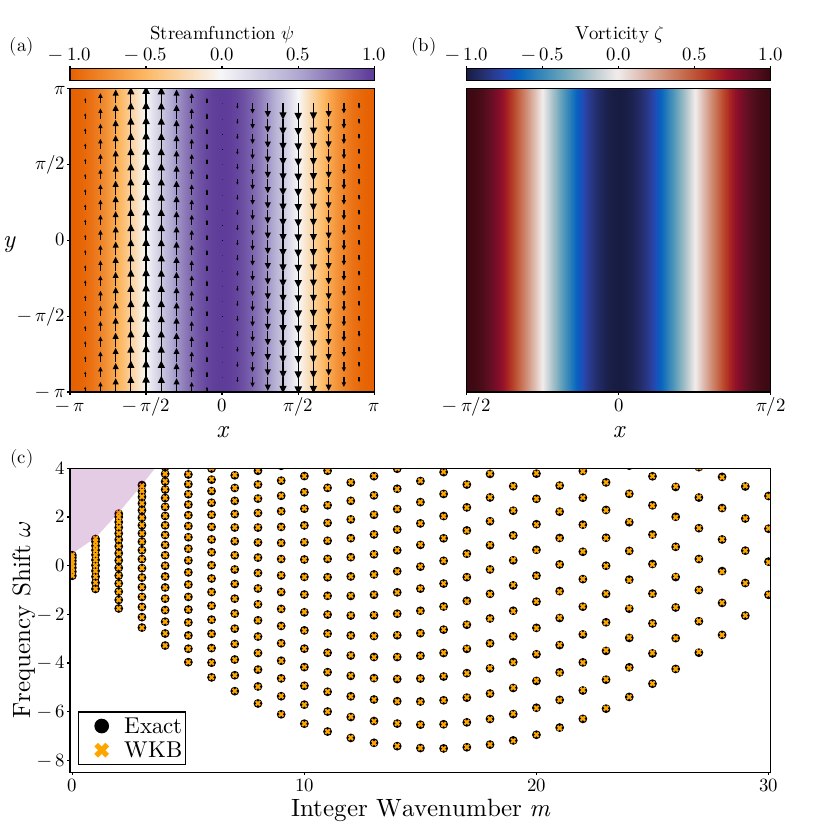}
  \caption{(a)~Streamfunction and flow vectors for the shear flow example. (b)~Vorticity showing the anticyclonic vorticity concentrated in the centre of the domain and cyclonic vorticity on the outside. (c)~Eigenvalues~$\omega$ as a function of the integer wavenumber $m$ for $\varepsilon=\frac{1}{4}$. The results from WKB theory (orange crosses) are shown along with the exact eigenvalues found from numerical solutions (black circles). The WKB results agree remarkably well with the numerical results although there are some spurious eigenvalues near the boundary between free and bound modes. The purple shading indicates the region where free modes exist, which are not shown here.}
  \label{fig:shear}
\end{figure}

We now consider a specific example of a parallel shear flow that varies sinusoidally in~$x$:
\begin{equation}
    \psi = \cos{x}.
\end{equation}
This shear flow has a region of anticyclonic vorticity at the centre of the domain and cyclonic vorticity centred on $x = \pm \upi$ (figure~\ref{fig:shear}a,b). This is a rare example in which the eigenvalue problem \eqref{eq:1D_shear} can be solved exactly using Mathieu functions \citep{balmforth1998}. The generally applicable WKB theory described above accurately predicts the eigenvalues, even for a modestly small~$\varepsilon = \frac{1}{4}$ (figure~\ref{fig:shear}c). We provide the analytical solutions to the WKB integrals in Appendix~\ref{app_analytical}. We also note that the symmetry of the problem means that a uniform wind forcing only projects onto modes with even~$n$.

\begin{figure}
    \centering
    \includegraphics{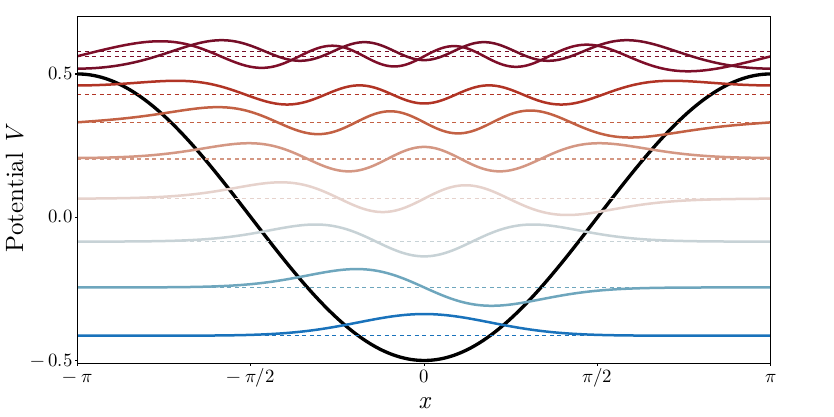}
    \caption{The potential $V = \frac{\zeta}{2}$ (black line) of the parallel shear flow for the $m=0$ mode. The dashed lines show the level of each eigenvalue~$\omega$ for $\varepsilon = \frac{1}{4}$. The solid coloured lines represent a scaled representation of each eigenfunction corresponding to a given eigenvalue, as identified by the colours.}
    \label{fig:shear_evecs}
\end{figure}

For $m = 0$, the eigenmodes are shaped by the potential~$V = \frac{\zeta}{2}$ (figure~\ref{fig:shear_evecs}). Where $\omega > V$, $S_0$ is imaginary and the solutions are oscillatory; where $\omega < V$, $S_0$ is real and the solutions are decaying (figure~\ref{fig:shear_evecs}). Near the anticylonic centre of the flow, the potential is at its lowest and all the modes are oscillatory. Moving further out into the cyclonic region, more and more of the modes become evanescent. The proportionality of the potential to the vorticity~$\zeta$ leads to trapping of NIW in anticyclones. The trapping arises from the dephasing of the modes that make up the initial condition. This is analogous to the argument in \citet{gill1984behavior} regarding the vertical propagation of NIWs due to the $\beta$-effect. 

\subsection{Axisymmetric flow}

We now consider a streamfunction with axial symmetry, such that $\psi = \psi(r)$, where $r$ is the radial distance from the origin. \citet{llewellyn1999near} studied NIWs with azimuthal wavenumber zero in an axisymmetric vortex and provided asymptotic expressions for the frequency of the lowest radial mode. \citet{kafiabad2021interaction} studied a similar case but also considered the impact of NIWs back on the vortex. Using WKB theory, we consider NIWs with an arbitrary azimuthal wavenumber and provide a transcendental equation that can be solved for their frequency as for the parallel shear flows above.

We make the ansatz $\phi=A(r)e^{im\theta}$, where $\theta$ is the azimuthal angle and again drop the mode label. In polar coordinates, \eqref{eqn:generaleig} then reduces to
\begin{equation}
  -\frac{\varepsilon^2}{2} \left( \dd{^2A}{r^2} + \frac{1}{r}\dd{A}{r} \right) + \left(\frac{\varepsilon^2 m^2}{2r^2} + \frac{mv}{r} + \frac{\zeta}{2} \right) A = \omega A,
\end{equation}
where $v = \partial_r \psi$ denotes the azimuthal velocity. There are some subtleties involved in applying WKB theory to this equation. For modes with $m>0$, the potential diverges at the origin. This issue has long been noted in the quantum mechanics literature and can be addressed by performing a so-called Langer transform on the equation. For $m=0$, there is no divergence of the potential, but there is a phase shift at the origin. As pointed out by \citet{berry1973semiclassical}, both cases turn out to give the same quantisation condition:
\begin{equation}
  \frac{\sqrt{2}}{\varepsilon}\int_{r_0}^{r_1}\sqrt{\omega-V(r)} \, \d r = \left(n+\frac{1}{2}\right)\upi, \quad \text{with} \quad n = 0, 1, 2, \dots,
\end{equation}
where the potential is
\begin{equation}
  V(r) = \frac{\varepsilon^2m^2}{2r^2}+ \frac{mv}{r} + \frac{\zeta}{2}.
\end{equation}
If $m > 0$, the integration bounds $r_0$ and $r_1$ are the two zeros of the integrand; if $m = 0$, $r_0 = 0$ and $r_1$ is the one zero of the integrand. As in the case of a parallel shear flow, this expression is uniformly valid in the sense that it works for both $m \sim O(1)$ and $m \sim O(\varepsilon^{-2})$. These again correspond to anisotropic and isotropic modes, respectively, with refraction, advection, and dispersion along and across streamlines playing the same roles as before. The only difference is that the streamlines are now circular.

\begin{figure}
    \centering
    \includegraphics{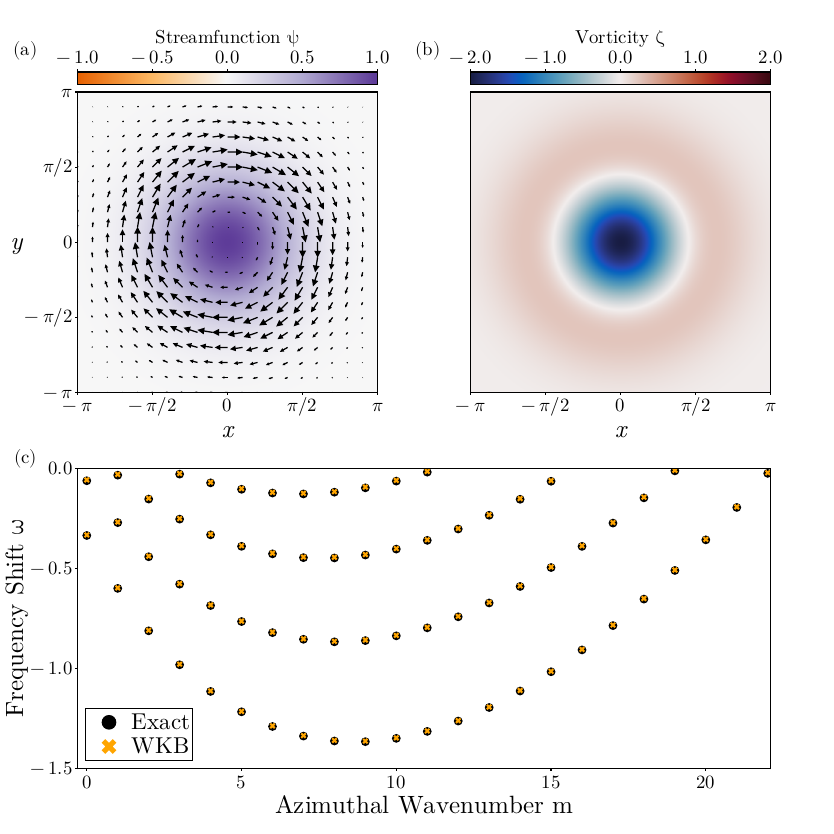}
    \caption{(a)~Streamfunction and flow vectors for the axisymmetric flow example. (b)~Vorticity field showing the anticyclonic vorticity concentrated in the centre of the domain which is flanked by a halo of cyclonic vorticity. (c)~Eigenvalues~$\omega$ as a function of azimuthal wavenumber~$m$ for~$\varepsilon=\frac{1}{4}$. The results from WKB theory (orange crosses) are shown along with the exact eigenvalues found from numerical solutions (black circles). The WKB approximation agrees remarkably well with the numerical results.}
    \label{fig:axi_a}
\end{figure}

\begin{figure}
    \centering
    \includegraphics{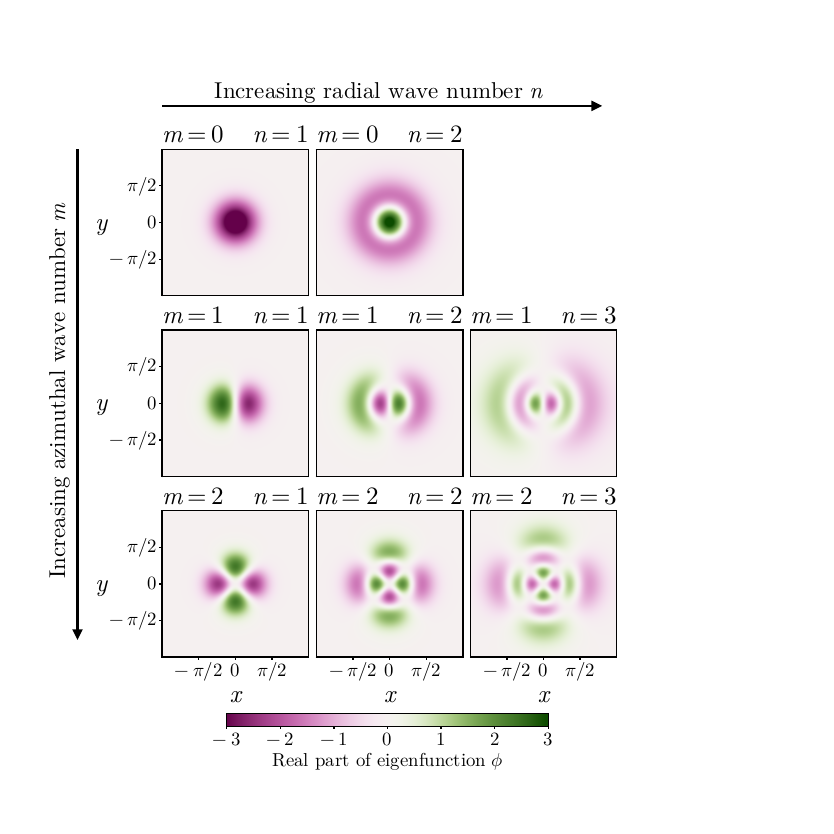}
    \caption{Real part of the eigenfunctions for the axisymmetric flow example with $\varepsilon = \frac{1}{4}$. The radial wavenumber~$n$ increases from left to right and corresponds to an increasing number of nodes in the radial direction. The azimuthal wavenumber increases from top to bottom and corresponds to an increasing number of nodes in the radial direction.}
    \label{fig:axi_evecs}
\end{figure}

We consider the concrete example of an isolated Gaussian vortex on an infinite domain:
\begin{equation}
  \psi(r)=e^{-\frac{r^2}{4}}.
\end{equation}
This corresponds to an anticyclone in the centre of the domain that is surrounded by a halo of cyclonic vorticity (figure~\ref{fig:axi_a}a,b). Again the WKB calculation yields eigenvalues that agree extremely well with the exact eigenvalues (figure~\ref{fig:axi_a}c). The structure of the first few modes is shown in figure~\ref{fig:axi_evecs}. For $m=0$, the modes are concentrated in the anticyclone. For $m>0$, there is a repulsion from the very centre of the anticyclone due to the advection and dispersion terms in $V(r)$. This repulsion increases with $m$, but the modes remain primarily concentrated in the region of anticyclonic vorticity. These modes are anisotropic, with more rapid variation in the radial (cross-streamline) than azimuthal (along-streamline) direction.

\subsection{General case}

Based on the intuition gained above, we wish to construct a uniformly valid asymptotic expansion for a general two-dimensional background flow. In analogy with the isotropic scaling, we begin by assuming a solution of the form
\begin{equation}
  \phi(x,y) = \exp\left[{\frac{1}{\varepsilon^2}\sum_{j=0}^\infty \varepsilon^{2j}S_{j}(x,y)}\right],
\end{equation}
again dropping the mode label. Substituting this into~\eqref{eqn:generaleig} yields
\begin{equation}
    -\frac{1}{2\varepsilon^2}\left|\sum_{j=0}^\infty \varepsilon^{2j}\nabla S_j\right|^2 - \frac{1}{2}\sum_{j=0}^\infty\varepsilon^{2j}\nabla^2S_j - \frac{i}{\varepsilon^2}\sum_{j=0}^\infty\varepsilon^{2j}\J(\psi,S_j) + \frac{\zeta}{2} = \omega.
\end{equation}
Assuming $\omega \sim O(\varepsilon^{-2})$ and collecting leading-order terms, we obtain
\begin{equation}
  -\frac{1}{2}|\nabla S_0|^2 - i\J(\psi,S_0) = \varepsilon^2 \omega.
  \label{eq:eps_-2}
\end{equation}
In the simple examples discussed above, we obtained a uniformly valid approximation by retaining the higher-order refraction term in the leading-order equation arising from an isotropic scaling. We do so again here:
\begin{equation}
  -\frac{1}{2}|\nabla S_0|^2 - i\J(\psi,S_0) +\frac{\varepsilon^2\zeta}{2} = \varepsilon^2 \omega.
  \label{eq:eps_-2_2}
\end{equation}
We anticipate that the order of these terms again changes for anisotropic modes. If the phase varies slowly along streamlines, the advection term is reduced by a factor~$O(\varepsilon^2)$, and cross-streamline dispersion, acting on spatial variations on a scale of~$O(\varepsilon)$ rather than $O(\varepsilon^2)$, will attain the same order, whereas along-streamline dispersion becomes negligible. The equation~\eqref{eq:eps_-2_2} can therefore capture both isotropic and anisotropic modes.

We now introduce the wavenumber vector~$\vec{k}$ by writing $\varepsilon^{-2} \partial S_0 / \partial \vec{x} = i \vec{k}$. The equation~\eqref{eq:eps_-2_2} can be solved using the method of characteristics:
\begin{equation}
  \dd{\vec{x}}{\tau} = \varepsilon^2 \vec{k} + \vec{u}, \qquad \dd{\vec{k}}{\tau} = -\pp{}{\vec{x}} \left( \vec{u}\vcdot\vec{k} + \frac{\zeta}{2} \right), \qquad \omega = \frac{\varepsilon^2 |\vec{k}|^2}{2} + \vec{u} \vcdot \vec{k} + \frac{\zeta}{2}.
\label{eq:WD}
\end{equation}
These are the non-dimensionalised ray-tracing equations of \citet{kunze1985}. We further elaborate on this connection between YBJ and \citet{kunze1985} below.

\begin{figure}
    \centering
    \includegraphics{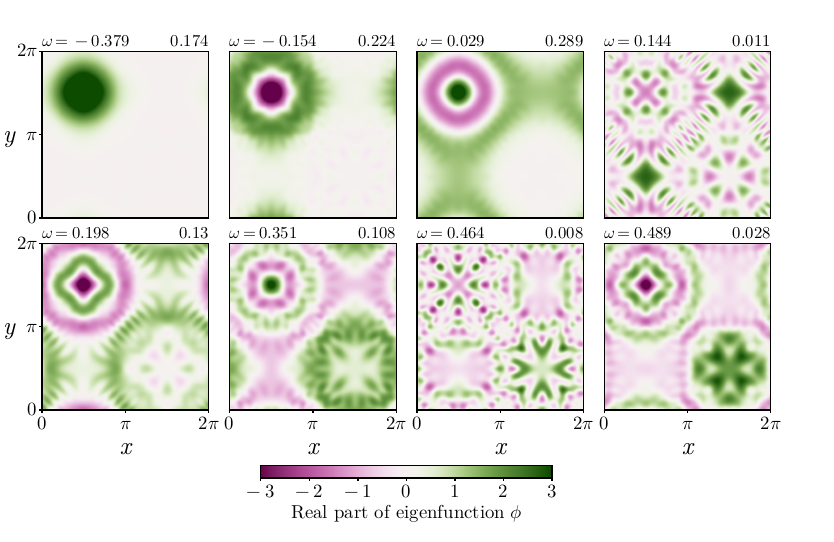}
    \caption{Real part of the eigenfunctions of the dipole flow with~$\varepsilon=\frac{1}{4}$. Together, these eight eigenfunctions represent over 97\% of the energy excited by a uniform impulsive forcing. They are the eight modes with the strongest projection and are then ordered by eigenvalue~$\omega$. The eigenvalues are shown in the top left corner and the projections of a uniform forcing onto the eigenfunction (energy fraction) are shown in the top right corner.}
    \label{fig:2D}
\end{figure}

Numerical solutions for the dipole flow show that the majority of a uniform forcing projects onto anisotropic modes that show little structure along streamlines and vary more rapidly across streamlines (figure~\ref{fig:2D}). With $\varepsilon = \frac{1}{4}$ there is also some projection onto modes that show more characteristics of isotropic phase variations. The variations are more rapid, as emerges from the isotropic scaling discussed above.

Finally, we show how approximations to the eigenvalues can be obtained in the weak-dispersion limit when the flow problem is not separable, as it was in the cases of a parallel shear flow or axisymmetric flow. To this end, we utilise results from the quantum mechanics literature. Recall that the YBJ equation is equivalent to the Schr\"odinger equation, with the YBJ operator
\begin{equation}
  H = -\frac{\varepsilon^2}{2} \nabla^2 - i \vec{u} \cdot \nabla + \frac{\zeta}{2}
\end{equation}
playing the role of the Hamiltonian. The weak-dispersion limit corresponds to the classical limit of the equivalent quantum system, and the ray-tracing equations are the analogue of the classical Hamiltonian dynamics. The classical Hamiltonian is obtained from~$H$ by making the substitution $\nabla \rightarrow i\vec{k}$, yielding the dispersion relation in~\eqref{eq:WD}. The Hamiltonian dynamics are then
\begin{equation}
  \dd{\vec{x}}{\tau} = \pp{\omega}{\vec{k}} \quad \text{and} \quad \dd{\vec{k}}{\tau} = -\pp{\omega}{\vec{x}},
\end{equation}
the ray-tracing equations stated in \eqref{eq:WD}. The connection with the Schr\"odinger equation is most easily seen in the Hamilton--Jacobi description of classical mechanics \citep[e.g.,][]{sakurai2020modern,buhler2006brief}.

\begin{figure}
  \centering
  \includegraphics{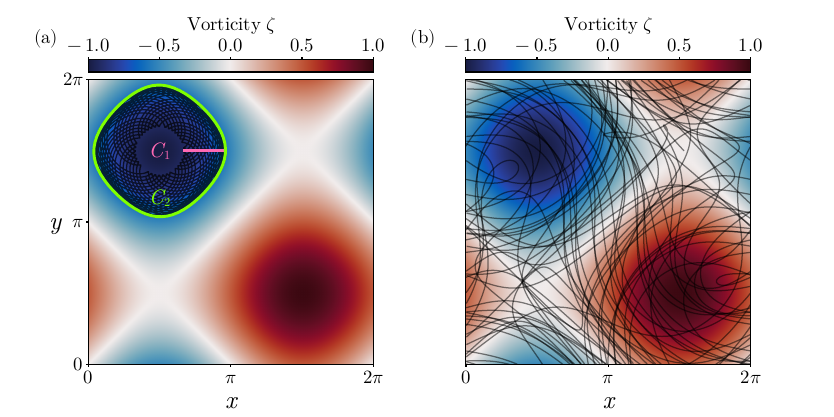}
  \caption{(a)~Example of a trajectory tracing out an invariant torus for the dipole case. This torus corresponds to $n=2$, $m=0$ for $\varepsilon = \frac{1}{4}$. The background colours show the vorticity field. The black line shows a finite-time trajectory on the torus. The green and magenta lines represent a choice for the two invariant curves on the torus. They are independent because no continuous deformation of one can transform it into the other. (b)~Different initial conditions result in different trajectories. This example is not bound to an invariant torus but is instead an example of a chaotic trajectory.}
    \label{fig:torus}
\end{figure}

The quantisation conditions derived above for separable problems, from which we obtained good approximations of the frequency shifts~$\omega$, can be generalised to some extent to non-separable problems like the dipole flow (figure~\ref{fig:dipole}). This semi-classical analysis of a quantum system was developed by \citet{einstein1917quantum}, \citet{brillouin1926remarques}, and \cite{keller1958corrected}, extending the Bohr--Sommerfeld quantum theory. The resulting approach is referred to as the EBK method \citep[see also][]{keller1985semiclassical,berry1972semiclassical,percival1977semiclassical}. The starting point is that the rays (classical trajectories in the quantum problem), being constrained by the invariant~$\omega$ (energy in the quantum problem), trace out invariant tori in the phase space spanned by $\vec{x}$ and~$\vec{k}$. A ray starting on such a torus will remain on it forever. The quantisation condition selects invariant tori that correspond to allowed bound states by insisting that phase increments along closed loops on the invariant torus integrate to multiples of~$2\upi$. Recalling that $\varepsilon^{-2} S_0 = i \vec{k}$, so $\vec{k}$ is the spatial gradient of the phase, and $\vec{k} \vcdot \d \vec{x}$ is a phase increment, the quantisation conditions read
\begin{equation}
  \oint_{\mathcal{C}_1}\vec{k}\vcdot \, \d\vec{x} = 2\upi\left(n+\frac{1}{2}\right), \qquad \oint_{\mathcal{C}_2}\vec{k}\vcdot \, \d\vec{x} = 2\upi m,
  \label{eq:EBK_quant}
\end{equation}
where $n$ and $m$ are integers. The contours $\mathcal{C}_1$ and $\mathcal{C}_2$ are topologically independent closed curves on the invariant torus (figure~\ref{fig:torus}a). In our example, the curve~$\mathcal{C}_1$ passes through the hole of the phase space torus, whereas the curve~$\mathcal{C}_2$ goes around the hole. The two curves are independent in the sense that neither one can be continuously deformed into the other. There is a half-integer phase shift in the quantisation condition arising from the integral along curve~$\mathcal{C}_1$ because this curve passes through two caustics, the generalisation of a turning point, where additional phase shifts are incurred \citep{brillouin1926remarques,keller1958corrected,maslov1972theory}. The curve~$\mathcal{C}_2$ encounters no caustics. The integer wavenumbers $n$ and~$m$ correspond to the cross- and along-streamline variations, respectively. These EBK quantisation conditions are entirely analogous to the WKB quantisation conditions derived above for the separable parallel shear flow and axisymmetric flow.

\begin{figure}
    \centering
    \includegraphics{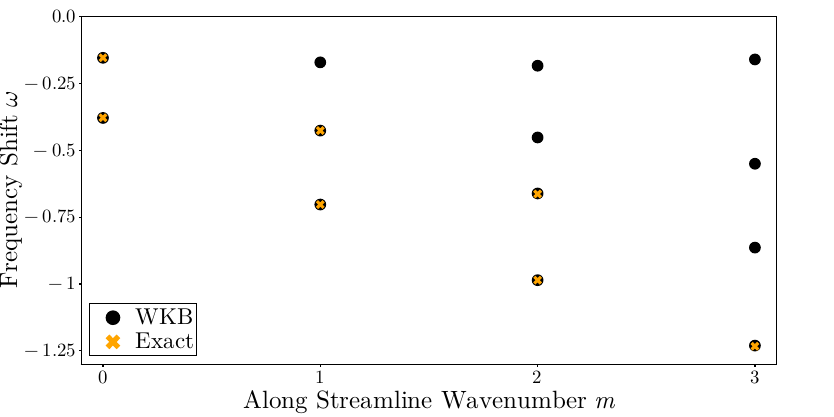}
    \caption{Numerical eigenvalues (black circles) and EBK eigenvalues (orange crosses) calculated for the dipole flow with $\varepsilon=\frac{1}{4}$. EBK calculations are only shown for the sufficiently confined modes where the invariant tori are easy to calculate. The EBK values agree with the numerical values to $\textit{O}(10^{-3})$.}
    \label{fig:EBK_modes}
\end{figure}

We apply the EBK quantisation to the dipole flow with $\varepsilon = \frac{1}{4}$. Our procedure closely follows \citet{percival1976vibrational}: we find the invariant tori satisfying the quantisation condition by writing the Hamiltonian equations in action--angle variables and employing Newton's method. See Appendix~\ref{app_EBK} for details. All eigenvalues calculated by this EBK method show excellent agreement with the numerical values (figure~\ref{fig:EBK_modes}).

As foretold by \citet{einstein1917quantum}, not all modes are accessible by the EBK approach. If the system is non-integrable, trajectories in phase space can become chaotic instead of tracing out an invariant torus (figure~\ref{fig:torus}b). States corresponding to such chaotic trajectories are not accessible by the EBK method. This ``quantum chaos'' has received much attention in the physics literature and has connections to random matrix theory \citep[e.g.,][]{gutzwiller1992quantum,stone2005einstein}. Methods exist to estimate eigenvalues as well as their statistics \citep[e.g.,][]{edelman2005random, edelman2007random}.  We do not pursue these issues any further here, in part because a uniform forcing projects most strongly onto the regular modes accessible with the EBK method (figure~\ref{fig:2D}). 

\section{Relation to the ray-tracing equations}

The previous section made clear that the ray-tracing equations of \citet{kunze1985} are closely related to the YBJ dynamics. In the same way that Hamiltonian dynamics emerge in the classical limit of the Schr\"odinger equation, the ray equations emerge in the weak-dispersion limit of the YBJ equation. YBJ criticised Kunze's assumption that the waves have a smaller spatial scale than the background flow, insisting that atmospheric forcing produces near-inertial waves at larger---not smaller---scales than mesoscale eddies, calling into question Kunze's ray-theoretical description in general. The analysis above clarifies that the spatial scale of the forcing is irrelevant. Instead, the scale on which dynamical modes vary determines whether WKB analysis can be applied, and this spatial scale is set by how strongly dispersive the waves are. An initially uniform wave field can be thought of as consisting of a superposition of several modes, all varying on a small scale but combining into a uniform field. The distinct frequencies~$\omega$ of these modes make them dephase over time and the superposition quickly exhibits the small scales of the modes.

Our analysis also provides some additional insight into the evolution of weakly dispersive NIWs. The isotropic and anisotropic scalings show that refraction is not always of leading-order importance. The refraction term is significant only for the anisotropic modes. For isotropic modes, the refraction term is asymptotically weak and the dispersion relation is dominated by advection and dispersion. A large-scale forcing, however, will project primarily onto the anisotropic modes, as can be seen in the specific solutions for the dipole case (figure~\ref{fig:2D}). More generally, the large values of the along-streamline wavenumber~$m$ in the isotropic case produce rapid variations that lead to strong cancellations when calculating the projection of a uniform forcing onto these modes. As such, only a weak projection can remain.

To help interpret observations from the NISKINe study, \citet{thomas2020} performed a simplified ray-tracing calculation which predicted a rapid strain-driven growth in the wavenumber that stood in stark contrast to the data. In this region of the North Atlantic, the waves are weakly dispersive \citep{thomas2024why}, so one may worry that this result contradicts our conclusion that ray-tracing can be deployed gainfully in the weak-dispersion regime. \citet{thomas2020} approximated the full wavevector evolution by assuming a uniform and time-independent vorticity gradient as well as a strain field with strain rate~$\alpha$ and its principal axis aligned with the vorticity gradient. In that setup, the wavenumber component~$k_\perp$ that is aligned with the vorticity gradient, \textit{i.e.}, perpendicular to vorticity contours, evolves according to
\begin{equation}
  \dd{k_\perp}{\tau} = -\frac{|\nabla \zeta|}{2} + \alpha k_\perp, \quad \text{so} \quad k_\perp = -\frac{|\nabla \zeta|}{2\alpha} \left( e^{\alpha \tau} - 1 \right)
\end{equation}
if $k_\perp = 0$ at time $\tau = 0$, approximating large-scale wind forcing. The exponential growth predicted by this equation does not match the data. Our analysis suggests, however, that a large-scale forcing primarily excites modes whose phase is aligned with streamlines. In this configuration, the strain is ineffective, and the initial wavenumber evolution is dominated by refraction:
\begin{equation}
  \dd{k_\perp}{\tau} = -\frac{|\nabla \zeta|}{2}, \quad \text{so} \quad k_\perp = -\frac{|\nabla \zeta|}{2} \tau.
\end{equation}
This recovers the \citet{asselin2020refraction} solution that \citet{thomas2020} showed roughly matches the data. Our analysis therefore suggests that it was not ray-tracing \textit{per se} that caused the mismatch with the data but the assumptions that went into the simplified solution.

\citet{kunze1985} considered three-dimensional ray-tracing which allows for both baroclinic structure in the mean flow and a vertical wavenumber for the NIWs that corresponds to propagation of the waves in the vertical. In this paper we have simplified to a barotropic mean flow and considered the propagation of a single baroclinic mode such that the problem reduces to two-dimensional ray-tracing. Exploring how the three-dimensional ray-tracing is related to the full YBJ equation that also allows for baroclinicity in the background flow is left to future work.

\section{Near-inertial wind work}
One may speculate that the frequency shifts in the weak-dispersion limit could impact the energy input into NIWs by the winds. To study this, we need to consider a forced version of the YBJ equation. So far we have focused on the problem with a horizontally uniform initial condition. This was to represent the NIW field excited by the passage of a large-scale atmospheric storm, and we studied the evolution of this NIW field in the absence of any further forcing. Real NIWs, in contrast, are continually forced by the winds, which we now represent by including a horizontally uniform forcing term in the modal YBJ equation:
\begin{equation}
  \d a_t = \left( -i \omega a_t + F_t e^{ift} \right) \d t,
  \label{eq:modal_forced}
\end{equation}
where $a_t$ denotes the modal amplitude at time~$t$ and $F_t$ the wind forcing projected onto the mode under consideration. We suppress the mode index~$\vec{\mu}$ for now, but keep in mind that this equation must be solved for each mode. Note that we have re-dimensionalised the equation here. The factor of $e^{ift}$ back-rotates the forcing to match the back-rotated description of the NIW evolution by the YBJ equation. To proceed, we describe the wind by an Ornstein--Uhlenbeck process which satisfies
\begin{equation}
  \d F_t = -c F_t \, \d t + \sigma \, \d W_t
\end{equation}
where $c^{-1}$ is the decorrelation time scale of the wind forcing, $\sigma$ is the amplitude of the stochastic excitation and $\d W_t$ is a Wiener process. The power spectrum of the process~$F$ is
\begin{equation}
  S(\omega) = \frac{2}{\upi} \frac{c}{c^2 + \omega^2}.
\end{equation}
For $\omega \gg c$ the power falls of with frequency as $\omega^{-2}$, \textit{i.e.}, the spectrum is red. We find that this is a good model of the power spectrum of the wind stress from reanalysis, especially over the ocean (see Appendix~\ref{app_c} for more details). 

We consider the system spun up from $t=-\infty$, such that it has statistically equilibrated for all~$t$. This results in the formal solution for the forcing
\begin{equation}
  F_t = \sigma \int_{-\infty}^t e^{-c(t-t')} \, \d W_{t'}, \label{eq:forcing_solution}
\end{equation}
and the formal solution for the mode amplitude~$a$ is given by
\begin{equation}
  a_t = e^{-i\omega t} \int_{-\infty}^t e^{i(f + \omega) t'} F_{t'} \, \d t'. 
\end{equation}
The NIW kinetic-energy equation can be obtained in the usual way by multiplying \eqref{eq:modal_forced} with $a_t^*$ and adding the complex conjugate. This is allowed because it is the integral of a Wiener process that appears in  \eqref{eq:modal_forced}, and not the Wiener process itself. The wind work $\Gamma_t$ arises as
\begin{equation}
  \Gamma_t = \frac{1}{2}\left(a_t^* e^{ift} F_t + \text{c.c.}\right).
\end{equation}
We are interested in the average of $\Gamma_t$ over an ensemble of many realisations of the wind-forcing. Let $\langle \, \cdot \, \rangle$ denote such the ensemble average. Hence, the ensemble average wind work is
\begin{equation}
  \langle \Gamma_t \rangle = \frac{1}{2}\left(e^{-i(f + \omega) t} \int_{-\infty}^t e^{-i(f+\omega) t'} \langle F^*_{t'} F_t \rangle \, \d t'+\text{c.c.}\right).
\end{equation}
The covariance function of the Ornstein--Uhlenbeck process~$F$ is
\begin{equation}
  \langle F^*_{t'} F_t \rangle = \frac{\sigma^2}{2c}e^{-c|t-t'|},
\end{equation}
so the ensemble average of $\Gamma_t$ reduces to
\begin{equation}
  \langle \Gamma_t \rangle = \frac{1}{2}\frac{\sigma^2}{c^2+(f+\omega)^2}.
\end{equation}
As expected, given the initialisation at $t = -\infty$, the power input is independent of time~$t$. From this expression, we can furthermore see that $\langle \Gamma_t \rangle$ is smaller for $\omega > 0$ than for $\omega < 0$. This is because the wind forcing has more power at low frequencies.

We now define $Q$ as the ratio between the equilibrium wind work in the presence of a mesoscale eddy field to the equivalent wind work in the absence of mesoscale eddies. Without mesoscale eddies, $\psi=0$ and hence there is no advection or refraction. Furthermore, if we assume the waves are generated by large-scale storms such that we approximate the forcing as horizontally uniform, there is no process to generate horizontal structure in the waves. The Laplacian of a constant wave field is zero and so the dispersion term also drops out of the YBJ equation. In this case there are no frequency shifts and $\omega = 0$ for the uniform mode excited by the wind. The wind work is simply
\begin{equation}
  \langle \Gamma_t \rangle = \frac{1}{2}\frac{\sigma^2}{c^2+f^2}.
\end{equation}
We calculate $Q$ as a weighted sum of the ratio over individual modes where the weighting is given by the projection $F_{\vec{\mu}}$ of the forcing onto a given mode $\vec{\mu}$:
\begin{equation}
  Q = \sum_{\vec{\mu}} |F_{\vec{\mu}}|^2\frac{c^2+f^2}{c^2+(f+\omega_{\vec{\mu}})^2},
  \label{eq:Q_infinty}
\end{equation}
where we restored the subscripts for the modes. This expression depends on the dispersiveness~$\varepsilon$ through $F_{\vec{\mu}}$ and~$\omega_{\vec{\mu}}$.

\begin{figure}
    \centering
    \includegraphics{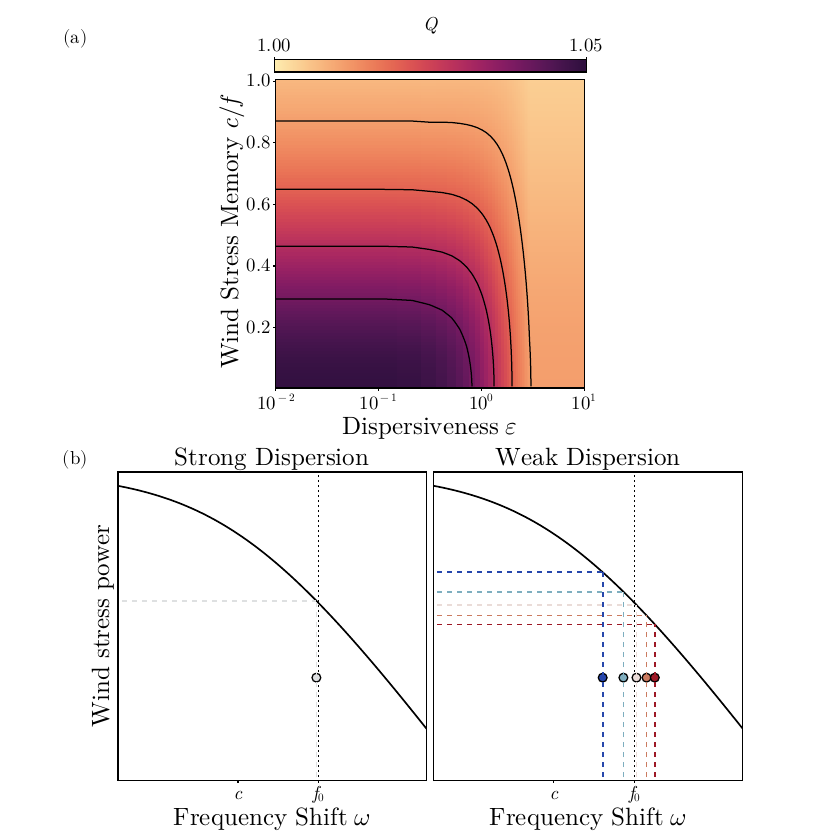}
    \caption{(a)~Ratio~$Q$ of NIW wind work in the presence of mesoscale eddies to that without as a function of the wave dispersiveness~$\varepsilon$ and the wind stress memory parameter $c/f$. Modulation of the NIW wind work by the mesoscale eddy field appears only for low $\varepsilon$ and $c/f$. These values correspond to a re-dimensionalisation of $\omega$ by $\Ro=0.5$. (b)~Schematic illustrating the enhancement of NIW wind work in the weak-dispersion regime. The solid black curve illustrates the wind stress power as a function of frequency on a log-log plot. At the inertial frequency, the power is rapidly falling off. The circles indicate the modes that a uniform initial condition projects onto. In the strong-dispersion case, the forcing projects onto a single mode with a small frequency shift. In the weak-dispersion case, the forcing projects onto a wide variety of modes with large frequency shifts.}
    \label{fig:Q_2d}
\end{figure}

Modulation of the NIW wind work by mesoscale eddies occurs only for $\varepsilon\lesssim 1$. Using the dipole flow as an example, we calculate $Q$ from \eqref{eq:Q_infinty} as a function of $c$ and~$\varepsilon$ (figure~\ref{fig:Q_2d}). For large~$\varepsilon$, $Q$ quickly approaches unity regardless of the value of $c$. For small $\varepsilon$, the contours of $Q$ become horizontal and there is little dependence of $Q$ on~$\varepsilon$. The dependence is primarily on $c$ with a lower value of $c$ resulting in a higher value of $Q$, \textit{i.e.}, a more substantial enhancement of the wind work.

Our framework provides physical motivation for why mesoscale eddies can modulate the wind work in the weak-dispersion case. Assuming $c \ll f$, which is generally the case for the wind stress over the ocean, we see that the inertial frequency~$f$ is in the $\omega^{-2}$ part of the wind power spectrum. Any process that shifts the frequency of NIWs will modulate the wind power felt by the waves. Because the wind power spectrum falls off like $\omega^{-2}$, a shift to lower frequencies will raise the wind power felt by the waves, and a shift to higher frequencies will lower it. This is the essence of \eqref{eq:Q_infinty}. As we have shown above, frequency shifts are small in the strong-dispersion limit and so the waves should feel similar wind power regardless of the presence of mesoscale eddies. As such, $Q$ is close to unity in the strong-dispersion limit. In the weak-dispersion limit, in contrast, there can be significant frequency shifts. A uniform forcing will project onto many modes with a range of frequency shifts. Due to the curvature of the wind power spectrum, going like $\omega^{-2}$, the fractional increase in power for negative frequency shifts will be greater than the fractional decrease in power for positive frequency shifts of the same magnitude. As a result, there will be a net increase in NIW wind work when summing over all modes (see figure~\ref{fig:Q_2d}b for a schematic). The question remains whether this will be an appreciable effect in the ocean.

We estimate~$Q$ from observations. For each location in the ocean, we estimate~$\varepsilon$ from the deformation radius and satellite altimetry observations of the eddy field (see Appendix~\ref{app_eps}), and we estimate $c$ from atmospheric reanalysis (see Appendix~\ref{app_c}). We calculate the modes of the dipole flow for a range of~$\varepsilon$, which gives us $\omega_{\vec{\mu}}$ and $|F_{\vec{\mu}}|^2$, and we re-dimensionalise $\omega_{\vec{\mu}}$ using the Rossby number $\Ro=\zeta/f$ calculated from satellite altimetry. We use the spatial structure of the vortex dipole as a stand-in for the real eddy structure. This provides an (admittedly crude) estimate of the combined effect of an anti-cyclone and a cyclone. We calculate $Q$ by using~\eqref{eq:Q_infinty} and then interpolating onto the correct~$\varepsilon$.

Our estimate reveals that deviations of $Q$ from unity are weak; at most~5\%. This effect is entirely concentrated in the western boundary current regions. This is because the dimensional frequency shift scales with $\Ro$. Over most of the ocean $\Ro$ is far too weak to produce any modulation of the NIW wind work. While this mechanism may be important for individual NIW events, it is clear that on average there is not a significant modulation of the NIW wind work by mesoscale eddies. The maximum modulation of~5\% is significantly smaller than current uncertainties in the NIW wind-work \citep{alford2020}. That being said, our approximation of the wind-stress as an Ornstein--Uhlenbeck process is crude. The real forcing is dominated by intermittent atmospheric cyclones. 

\section{Limitations of the model}

The scaling assumptions of the original YBJ equation place limits on~$\varepsilon$, which should be kept in mind in the context of the asymptotic expansions performed above. The Rossby number is $\Ro = U/fL$, where $U$ is a scaling for the background flow, and the Burger number is $\Bu = \lambda^2/L^2$. With these definitions, $\varepsilon^2 = \Bu / \Ro$. Both $\Ro$ and $\Bu$ are required to be small but YBJ go further and make the scaling assumption $\Ro \sim \Bu$, which is equivalent to $\varepsilon^2 \sim 1$. See \citet{thomas2017near} for a discussion of other dispersion regimes.

The asymptotic analyses performed above technically violate the original scaling assumption made by YBJ, but they still offer insight. The parameter~$\varepsilon$ does not need to be far into either the strong or weak-dispersion regime for the solutions to exhibit characteristics of the asymptotic solutions. In the strong-dispersion limit, the modes for $\varepsilon=2$ strongly resemble the analytical strong-dispersion solutions. In the weak-dispersion limit, all of our examples show excellent agreement between asymptotic theory and numerical solutions for $\varepsilon = \frac{1}{4}$. This behaviour is typical of WKB solutions, which often work well beyond where they have any right to be accurate.

\citet{thomas2017near} conducted a detailed study of the evolution of NIWs in different scaling regimes. They considered a \enquote{very weak--dispersion regime} where $\Bu\sim\Ro^2$ which is equivalent to $\varepsilon\sim\Ro$. An additional term arises compared to the YBJ equation, but they found the YBJ equation to still work well in simulations. They also consider a \enquote{strong-dispersion regime} where $\Bu\sim1$. In this regime they find a leading-order uniform NIW solution, but also with the excitation of super-inertial frequencies that are not captured by YBJ. The frequency shift of the uniform mode is as predicted by YBJ.

We can further assess the validity of YBJ's scaling assumption using our observational estimates of~$\varepsilon$. For the first four baroclinic modes, most of the ocean is in the regime of $\varepsilon\sim1$ (Fig.~\ref{fig:epsilon}). For higher baroclinic modes, $\varepsilon$ will continue to decrease proportionally to $\lambda^{-1}$. For high-enough baroclinic mode number, the $\varepsilon^2\sim1$ requirement will therefore be broken. For many regions, however, these high baroclinic modes may not be strongly excited by the wind forcing \citep[e.g.,][]{balmforth1998}, although this will not be true universally \citep[e.g.,][]{thomas2024why}.

Throughout this paper, we have dealt with the case in which the background flow does not evolve. In the ray-tracing framework, the background flow could be allowed to evolve. The Hamiltonian operator would be time-dependent, but the equations can still be integrated along rays. For our analysis of eigenmodes to be applicable to the time-dependent case, however, the evolution of the background flow should be adiabatic, \textit{i.e.}, it should be slow compared to the wave evolution. The time for eigenmodes to de-phase depends on the difference between their frequencies. In the strong-dispersion case, the frequency difference between the leading-order eigenmode and the higher eigenmodes is $O(\varepsilon^2)$, meaning that the time to de-phase should be small relative to the timescale for evolution of the background flow. In the weak-dispersion limit, the eigenvalues become ever-closely packed, meaning the timescale for dephasing can become large. For the adiabatic assumption to hold, an invariant torus should deform much more slowly than the time it takes a particle to traverse the torus. If the time taken to traverse the torus is given by the advective timescale then these two timescales are formally the same order. The adiabatic assumption will only hold if there is a symmetry which causes the torus to persist for a longer timescale. The dipole vortex is an extreme example of this where the tori never deform yet the advective timescale is finite. In the ocean, eddies often persist as coherent features for times much longer than the advective timescale. As such, we argue that the weak-dispersion results should continue to provide insight even in the time-dependent case.

\section{Conclusions}

In the YBJ framework, the evolution of NIWs in the presence of a mesoscale eddy field is governed by the wave dispersiveness~$\varepsilon = f \lambda^2 / \Psi$. The limit of $\varepsilon \gg 1$ corresponds to the strong-dispersion limit and $\varepsilon \ll 1$ corresponds to the weak-dispersion limit. Both of these limits are relevant for the ocean as the dispersiveness decreases with vertical mode number and the strength of mesoscale eddies.

The YBJ equation is a Schr\"odinger equation, with the YBJ operator playing the role of the Hamiltonian operator in quantum mechanics. As is conventional in quantum mechanics, the evolution of NIWs can be described using the eigenmodes of the YBJ operator and their eigenvalues, which determine the frequency shift away from the inertial frequency. Perturbation methods from quantum mechanics yield insight into the YBJ dynamics and its relationship to the ray-tracing equations of \citet{kunze1985}.

In the strong-dispersion regime~$\varepsilon \gg 1$, perturbation theory yields closed-form expressions for the NIW modes. To leading order, a spatially uniform forcing excites a spatially uniform NIW mode. This mode is modulated by an order~$\varepsilon^{-2}$ perturbation proportional to the streamfunction of the eddy field. The frequency shift is also of order~$\varepsilon^{-2}$ and proportional to the average kinetic energy of the eddies. Both of these results recover predictions by YBJ. The same approach also yields expressions for the modes that are not spatially uniform to leading order. The degeneracy of these modes at leading order is lifted at higher order, and the frequency shifts and spatial structures can be determined. Wind patterns associated with sharp atmospheric fronts may excite these modes more strongly than the uniform forcing assumed throughout this work \citep[e.g.,][]{thomas2017modifications}.

In the weak-dispersion regime~$\varepsilon \ll 1$, the YBJ equation is amenable to WKB analysis. In simple (separable) background flow geometries, this allows the straightforward calculation of eigenmodes and their frequency shifts, which are excellent approximations of the exact frequency shifts even for modestly small~$\varepsilon$. More generally, the weak-dispersion limit of the YBJ equation corresponds to the classical limit of quantum mechanics. The YBJ equation reduces to the ray equations of \citet{kunze1985}, the equivalent to the corresponding classical Hamiltonian dynamics. The semi-classical EBK analysis allows the calculation of frequency shift for non-separable background flows for the regular part of the spectrum, which again are in excellent agreement with the full shifts. The emergence of the ray equations in the classical limit furthermore suggests that they can be applied if dispersion is weak, whether or not the forcing has a large horizontal scale. The spatial-scale separation underlying the ray equations emerges because refraction quickly produces short-scale phase variations, initially unopposed by dispersion.

The frequency shift of NIW modes away from the inertial frequency implies that the NIW wind work can be modulated by mesoscale eddies. We quantified this using $Q$ which measures the ratio of the NIW wind work in the presence of mesoscale eddies to that without mesoscale eddies. This modulation arises due to the curvature of the wind power spectrum, which enhances the power input into modes with a shift to lower frequencies more than it suppresses the power input into modes with a shift to higher frequencies. On average, this effect is weak in the oceans, however, with the modulations always being less than~5\%.

\textit{Acknowledgements} The authors gratefully acknowledge support from NASA under grants 80NSSC22K1445 and 80NSSC23K0345, from NSF under grant OCE-1924354, and from the Simons Foundation Pivot Fellowship program.

\textit{Data Availability Statement} Code to numerically solve the 2D eigenvalue problem is available at \url{https://github.com/joernc/ybjmodes}. The SSH data is available from the E. U.’s Copernicus Marine Service at \url{https://doi.org/10.48670/moi-00148}.  The ERA5 reanalysis data is available from the Copernicus Climate Change Service (C3S) Climate Data Store at \url{https://doi.org/0.24381/cds.adbb2d47}. The ECCO density data is available from \url{https://doi.org/10.5067/ECG5D-ODE44}.

\textit{Declaration of Interests} The authors report no conflict of interest.

\appendix

\section{Calculating the wave dispersiveness}
\label{app_eps}

Here we describe the calculations used to estimate the wave dispersiveness~$\varepsilon = f \lambda^2 / \Psi$ from observations. At each location, we estimate the set of deformation radii~$\lambda$ from hydrography and the characteristic strength of the streamfunction~$\Psi$ from altimetry.

Following \citet{smith2007geography}, we calculate $\lambda$ by solving the baroclinic eigenvalue equation using finite differences. We perform this calculation using the climatology from the Estimating the Circulation and Climate of the Ocean (ECCO) state estimate version 4 release 4 \citep{fukumori2020synopsis,forget2015ecco}. We solve the baroclinic eigenvalue equation at each horizontal grid cell on the ECCO grid to obtain maps of the deformation radii. We display~$\varepsilon$ for the lowest four baroclinic modes only, for which the numerical approximation has a minimal effect.

To calculate $\Psi$, we use sea surface height (SSH) observations from the Data Unification and Altimeter Combination System's (DUACS) delayed-time (DT) 2018 release \cite{taburet2019duacs}. The SSH is provided at a (nominal) $(1/4)^\circ$ and daily resolution. We calculate a geostrophic streamfunction using $\psi = g\eta/f$, where $\eta$ is the SSH and $f$ is the (now latitude-dependent) Coriolis parameter. We take observations from 2007 to~2022 and estimate~$\Psi$ as the RMS $\psi$ over that period. Again we are assuming that the streamfunction is barotropic.

\section{Numerical solutions to the eigenvalue problem}
\label{app_num}

To numerically solve the eigenvalue problem~\eqref{eqn:generaleig}, we discretise the Hamiltonian operator~$H$ using finite differences. We use a fourth-order central difference scheme for the Laplacian operator in the dispersion term. For the advection term, we employ the fourth-order enstrophy-conserving scheme of \cite{arakawa1966computational}, which preserves the Hermitian nature of the operator and translates into energy conservation in this context. In the notation of \cite{arakawa1966computational}, we employ~$2J_1 - J_2$, where $J_1 = \frac{1}{3} (J^{++} + J^{+\times} + J^{\times+})$ and $J_2 = \frac{1}{3} (J^{\times \times} + J^{\times+} + J^{+\times})$. For the refraction term, we evaluate~$\zeta$ analytically at each point, although in general it could be calculated from the streamfunction using finite differences as well.

We use a spatial resolution of up to $1024 \times 1024$ points and solve for the lowest eigenvalues using Lanczos iteration. The resolution is chosen by checking the convergence of the eigenvalues. The number of eigenvalues solved for depends on the value of~$\varepsilon$, which controls how densely packed the eigenvalues are and thus how many must be computed to find all eigenmodes that a uniform forcing projects onto substantially. We ensure a large enough number of eigenvalues are computed by summing the square of the projection coefficients.

\section{Analytical solutions to shear flow WKB integrals}
\label{app_analytical}

Here we provide analytical solutions to the WKB problem with the sinusoidal shear flow. First we rewrite the potential as
\begin{equation}
    V(x)=A_m\cos{(x+\delta_m)}+\frac{\varepsilon^2m^2}{2},
\end{equation}
where $A_m=\sqrt{m^2+\frac{1}{4}}$ and $\tan{\delta_m}=-2m$. Assuming $m>0$ and that $\arctan$ corresponds to the principal value then it follows that $\delta_m=\upi+\arctan{\left(-2m\right)}$. Since the domain is periodic, we can consider any interval of length $2\upi$. For convenience we choose $[-\upi-\arctan(-2m),\upi-\arctan(-2m)]$. We can now make the change of variable $x'=x+\arctan(-2m)$. The transformed potential is
\begin{equation}
    V(x') = \frac{\varepsilon^2m^2}{2}-A_m\cos(x').
\end{equation}
With the potential in this form, the WKB integral \eqref{eq:S0} can be evaluated in terms of the elliptic integral of the second kind $E(\varphi|k^2)$
\begin{equation}
  S_0=\pm2\sqrt{2}i\varepsilon\sqrt{\omega-\frac{\varepsilon^2m^2}{2}+A_m}\;E\left(\frac{x'}{2}\Bigg|\frac{2A_m}{\omega - \frac{\varepsilon^2m^2}{2}+A_m}\right).
\end{equation}
We can obtain an equation for the eigenvalues from \eqref{eq:shear_eval_eqn}. Letting $x'_1$ denote the positive turning point given by $x'_1 = \upi-\arccos\left(\frac{\omega-\varepsilon^2m^2/2}{A_m}\right)$, we obtain
\begin{equation}
    E(\varphi|k^2)=\frac{\varepsilon\upi\left(n+\frac{1}{2}\right)}{4\sqrt{2\left(\omega-\frac{\varepsilon^2m^2}{2}+A_m\right)}},
\end{equation}
where $\varphi = \frac{x'_1}{2}$ and $k^2 = \frac{2A_m}{\omega-\varepsilon^2m^2/2+A_m}$. This is a transcendental equation that can be solved numerically for the eigenvalues $\omega$. 

The eigenvalues can be normalised by requiring
\begin{equation}
    \int_{-\upi}^{\upi}[\phi_n(x)]^2dx = 2\upi.
\end{equation}
Letting $C$ be the normalisation constant we obtain
\begin{equation}
    \frac{8C^2}{\sqrt{\omega-\frac{\varepsilon^2m^2}{2}+A_m}}F(\varphi|k^2)=2\upi,
\end{equation}
where $F(\varphi|k^2)$ is the elliptic integral of the first kind \citep[see][]{bender1999advanced}.

The projection of a uniform forcing onto a given mode with even symmetry about the bottom of the potential is
\begin{equation}
    a_n = \frac{1}{2\upi}\int_{-\upi}^\upi\phi_n(x)dx.
\end{equation}
This integral can be evaluated \citep[again see][]{bender1999advanced} as 
\begin{equation}
    |a_n|^2 = \frac{1}{\sqrt{A_m-\omega+\frac{\varepsilon^2m^2}{2}}}\frac{\varepsilon}{\sqrt{2}F(\varphi|k^2)}.
\end{equation}

\section{Further Details about the EBK Method}\label{app_EBK}

We principally follow \cite{percival1976vibrational} to calculate find the invariant tori satisfying quantisation conditions and the associated EBK predictions for the eigenvalues~$\omega$. We write the angle Hamilton equations, which are partial differential equations that describe the invariant torus:
\begin{align}
  \nu \pp{x}{\theta} + \mu \pp{x}{\varphi} &= \varepsilon^2 k + u \\
  \nu \pp{y}{\theta} + \mu \pp{y}{\varphi} &= \varepsilon^2 l + v \\
  \nu \pp{k}{\theta} + \mu \pp{k}{\varphi} &= -\left(k \pp{u}{x} + l \pp{v}{x} + \frac{1}{2} \pp{\zeta}{x} \right) \\
  \nu \pp{l}{\theta} + \mu \pp{l}{\varphi} &= -\left(k \pp{u}{y} + l \pp{v}{y} + \frac{1}{2} \pp{\zeta}{y} \right)
\end{align}
The quantization conditions can then be written as integrals over the angles $\theta$ and~$\varphi$:
\begin{equation}
  \int \left( k \pp{x}{\theta} + l \pp{y}{\theta} \right) \, \d \theta = 2\upi m, \qquad \int \left( k \pp{x}{\varphi} + l \pp{y}{\varphi} \right) \, \d \varphi = 2\upi(2n + 1 + m),
\end{equation}
The integration along~$\theta$ passes around the hole of the torus (like contour~$\mathcal{C}_2$ in figure~\ref{fig:torus}). The integration along~$\varphi$ passes through the hole twice and also around the hole once, so we double the radial phase increment $2\upi (n + \frac{1}{2})$ and add the azimuthal phase increment $2\upi m$ in the second quantization condition. We average these numerical integrals over the respective other coordinate to increase the accuracy.

We discretise the above equations by dividing the $[0, 2\upi]$ intervals that the angles $\theta$ and~$\varphi$ vary over using 64~point and approximate derivatives using an eighth-order finite difference scheme. We initialise the calculation with $\theta = -\frac{1}{2}$, $\varphi = \frac{1}{10}$, 
\begin{equation}
  x = \frac{1}{2} (1 + \cos \varphi) \cos \theta, \quad y = \frac{1}{2} (1 + \cos \varphi) \sin \theta, \quad k = -\varepsilon^{-1} \sin \varphi \cos \theta, \quad l = -\varepsilon^{-1} \sin \varphi \sin \theta
\end{equation}
for the $(n, m) = (0, 0)$ torus and apply Newton iteration to satisfy the above equations. We determine~$\omega$ by applying the dispersion relation at each point of the torus and averaging over all grid points. We then change the quantum numbers to other values and start the Newton iteration from the previously found torus, using iterations at intermediate values if needed.

\section{Estimating the decorrelation time of wind stress}
\label{app_c}

Here we describe the calculations used to estimate the decorrelation time~$c^{-1}$ of the wind stress. For the wind forcing, we use the European Centre for Medium-Range Weather Forecasting ERA-5 reanalysis \citep{hersbach2018}. For the calculations below, we use data from 2015 to~2020. At each grid cell, we use the 10~m zonal ($u_w$) and meridional ($v_w$) winds with hourly resolution. Following \citet{pollard1970} we convert this to a wind stress using a bulk aerodynamic drag formulation. The time series at each location is used to calculate a power spectrum of the wind stress. The decorrelation time scale is obtained by fitting the following model to the estimated spectrum:
\begin{equation}
    S(\omega) = \frac{A}{1+\left(\frac{\omega}{c}\right)^s},
\end{equation}
where $A$ and $s$ are additional fitted parameters that we do not use here. Over the ocean $s=2$ is a reasonable approximation which motivates our use of the Ornstein--Uhlenbeck process above.

\bibliographystyle{jfm}
\bibliography{bib}

\end{document}